\documentclass[%
apl,twocolumn,showpacs,superscriptaddress,letterpaper]{revtex4-2}

\usepackage{graphicx,amsfonts}% Include figure files
\usepackage{dcolumn}% Align table columns on decimal point
\usepackage{bm}% bold math
\makeatother
\usepackage[dvipsnames]{xcolor}
\definecolor{Mycolor2}{HTML}{C2185B}
\usepackage{hyperref}
\hypersetup{
	colorlinks=true,
	linkcolor=blue,
	filecolor=magenta,
	urlcolor=blue,
	citecolor= blue
}

\makeatother
\begin{document}

	\title{ Superconductivity, valence-skipping and topological crystalline metal in AgSnSe$_2$}
	
	\date{\today}
	\author{Shubham Patel}\email{spatelphy@iitkgp.ac.in}
	\affiliation{Department of Physics, Indian Institute of Technology, Kharagpur-721302, India}
	%\author{}
	\author{A Taraphder$^1$}\email{arghya@phy.iitkgp.ac.in}
	
	\begin{abstract}
		The recent suggestion of valence-skipping phenomenon driving a two-gap superconductivity in $Ag$-doped SnSe, by Kataria, \textit{et al.} [Phys. Rev. B 107, 174517 (2023)], has brought to the fore a long-standing issue once again. The absence of crystallographically inequivalent Sn cites corroborated by electronic properties of AgSnSe$_2$, calculated using first-principles density functional theory, however, does not appear to provide a strong support in favor of valence-skipping in this system. Interestingly, the signature of avoided band-crossings (with the inclusion of SOC) and non-zero \textit{mirror} Chern number ($n_{\mathcal{M}}$) confirm a non-trivial topology. The presence of mirror symmetry-protected surface states along the mirror planes indicates that AgSnSe$_2$ could be a potential candidate for topological crystalline metals (TCMs). Moreover, our calculation of electron-phonon coupling and anisotropic superconducting properties of AgSnSe$_2$, using Migdal-Eliashberg theory, gives a single-gap superconductivity with critical temperature $T_c \approx 7$K, consistent with the experimental value of $5$K. The interplay of topology and superconductivity in this three-dimensional material appears quite intriguing and it may provide new insights into the exploration of superconductivity and topology.
	\end{abstract}
	
	\maketitle
	\section{Introduction}
	
	Materials containing ions with valence-skipping (VS) tendency lead to the formation of negative-$U$ centers in certain materials. The resulting unretarded attractive interaction between the electrons may lead to enhanced, unconventional superconductivity due to strong charge fluctuations \cite{taraphder1991heavy,taraphder1995negative,taraphder1996exotic,varma1988missing}. $Ag$-doped SnSe is claimed to be a VS compound \cite{johnston1977superconducting} which shows superconductivity in the cubic rocksalt phase. There are controversies over the VS state in AgSnSe$_2$. In AgSnSe$_2$ there is a nominal valence of Sn, $+3$, which is expected to be skipped and the chemical formula of the compund could then be expressed as $\rm{(Ag^{1+})(Sn^{2+})_{0.5}(Sn^{4+})_{0.5}(Se^{2-})_{2}}$ as predicted by magnetic susceptibility \cite{johnston1977superconducting}, Sn M{\"o}ssbauer spectra in lead chalcogenieds \cite{nasredinov1999mossbauer} and a very recent muon spin rotation and relaxation measurement ($\rm{\mu SR}$) \cite{kataria2023superconducting}. On the other hand, there are contrary reports that do not find VS in this system. Consequently, they suggest anisotropic SC in this system \cite{zhi2013anomalous,naijo2020unusual}. The report by Naijo \textit{et al.} suggests that the unusual $+3$ valence in this system is likely because of geometrical constraint which prohibits a breathing distortion that could screen the on-site Coulomb repulsion \cite{naijo2020unusual}. This raises a question regarding the possible valence-skipping in AgSnSe$_2$. The superconducting transition temperature is also not very high, which is expected from an electronic mechanism of unconventional superconductivity. 
	
	However, negative-$U$ backed by the VS is not necessarily the only route to realise SC in these materials. Electron-phonon coupling (EPC) could play a vital role for the origin of SC here. The superconducting critical temperature of AgSnSe$_2$ is reported to be 4.93K \cite{johnston1977superconducting}, which is not very large and it appears that the electron-phonon mechanism or its anisotropic variant may be relevant in the context of superconductivity here. There are also disputes over the nature of superconducting gaps in the system. The recent $\rm{\mu SR}$ study \cite{kataria2023superconducting} suggests a two-gap SC while an earlier one does not find any trace of two-gap SC in AgSnSe$_2$ \cite{zhi2013anomalous}.
	
	It has been well establised by now that SnSe is a topological crystalline insulator (TCI) protected by crystal symmetry, with even number of Dirac cones, non-zero Chern number \cite{sun2013rocksalt,jin2017electronic}, and with efficient thermoelectric generation \cite{pletikosi2018band,kutorasinski2015electronic,zhao2016snse}. SnS and SnSe are both light TCIs discovered after the narrow band gap
	semiconductor SnTe \cite{hsieh2012topological,tanaka2012experimental} and the lead (Pb) based alloy Pb$_{1-x}$Sn$_x$Se$/$Te \cite{xu2012observation} with large spin-orbit coupling (SOC). TCIs are regarded topological insulators (TIs) with metallic surface states having quadratic band degeneracy protected by time-reversal and discrete rotational symmetry without the involvement of SOC \cite{fu2011topological}. Doping SnSe partially with $Ag$ (hole doping) induces superconductivity, but the topological behavior has not been discussed yet in this compound, neither experimentally nor theoretically.
	
	Motivated by the above discussions, we focus on the ongoing debates and embark on a comprehensive investigation of AgSnSe$_2$, aiming to elucidate its electronic, topological and superconducting nature through a multifaceted approach encompassing structural analysis by employing theoretical techniques based on first-principles density functional theory. From our analysis we do not find evidence for valence-skipping in this compound. Additionally, our investigation sheds light on the origin of topology in the system through hole doping using Ag, which shifts the Fermi energy (FE) into the valence band region and gives rise to a metallic Fermi surface, along with topologically protected band-inversion. This suggests a possible transition from normal metal to a novel class of topological materials, topological crystalline metals (TCMs). A TCI phase may be achieved in this system by tuning the chemical potential, electron doping or applying a gate-voltage in thin films. 

	As the system shows superconductivity, a possible EPC-induced SC is worked out. The phonon softening in the lowest acoustic mode originating from the Se atoms is identified as the possible source of SC in the system. Anisotropic superconducting properties are investigated by utilising Migdal-Eliashberg theory (MET) and AgSnSe$_2$ is likely to be a one-gap isotropic superconductor and not a two-gap one as proposed in one of the experimental studies \cite{kataria2023superconducting}.

	The paper is organized as follows: In Sec. \ref{sec:comp} we provide computational details which are followed by Sec. \ref{sec:elec} in which we discuss the crystal geometry and electronic band structure of AgSnSe$_2$. After identifying a band-crossing and band-inversion we further calculate the Berry curvature and topological Chern number in Sec. \ref{sec:topo}. Motivated by the experimental studies of SC in the system we investigate the anisotropic superconducting properties in Sec. \ref{sec:sc}. We conclude our findings in Sec. \ref{sec:conclusions}.
	
	\section{\label{sec:comp}Computational Details}
	The lattice dynamics, electronic structure and electron-phonon coupling (EPC) are calculated within the norm-conserving pseudopotentials for exchange-correlation functional \cite{troullier1991efficient}, as implemented in {\footnotesize QUANTUM ESPRESSO} (QE) package \cite{giannozzi2009quantum,giannozzi2017advanced,giannozzi2020quantum}, with a plane-wave energy cutoff of 70\,Ry and methfessel-paxton smearing of 0.01\,Ry. We calculated Berry curvature and Chern number numerically using the Hamiltonian we extracted from maximally localized Wannier functions (MLWFs) using  {\footnotesize WANNIER90} library \cite{marzari1997maximally,souza2001maximally,mostofi2008wannier90}. We Fourier transformed the real-space Hamiltonian into the $\mathbf{k}$-space and further calculated Berry curvature using a relation given by Eqn. \ref{eq:berry} in Sec. \ref{sec:topo}.
	
	The phonon dispersion is obtained by Fourier interpolation of the dynamical matrices computed using a $6\times6\times6$ \textbf{k}-point mesh and a $3\times3\times3$ \textbf{q}-point mesh. 
	The anisotropic superconducting properties are calculated using fine \textbf{k} and \textbf{q} grids of $60\times60\times60$ and $30\times30\times30$, respectively,  using MET as implemented in {\footnotesize EPW} code \cite{giustino2007electron,ponce2016epw,margine2013anisotropic}. For the Wannier interpolation in {\footnotesize EPW}, we used maximally localized Wannier functions to describe the electronic structure near the Fermi level. The Matsubara frequency cutoff is set to $0.2$ eV, which is 10 times larger than the upper limit of the phonon frequency in el-ph calculations. The mathematical and technical details of Migdal-Eliashberg calculations are described extensively by Allen \cite{allen1983theory}, Margine \cite{margine2013anisotropic} and Ponc{\'e} \cite{ponce2016epw} previously. Here, we concentrate on electronic, vibrational and superconducting properties. The Eliashberg electron-phonon spectral function $\alpha^2F(\omega)$ and the cumulative frequency dependence of EPC, $\lambda(\omega)$ can be calculate by 
	$$	\gamma_{qv} = 2\pi\omega_{qv}\sum_{nm}^{}\int_{BZ}^{} \frac{dk}{\Omega_{BZ}}|g_{mn,v}(k,q)|^2 $$
	\begin{equation}
		\label{eq:gamma}
		\times \delta(\epsilon_{nk}-\epsilon_F)\delta(\epsilon_{mk+q}-\epsilon_F).
	\end{equation}
	\begin{center}
		\label{eq:alpha2F}
		\begin{equation}
			\alpha^2F(\omega) = \frac{1}{2\pi N(E_F)}\sum_{qv}^{} \frac{\gamma_{qv}}{\omega_{qv}} \delta(\omega-\omega_{qv})		
		\end{equation}
	\end{center}
	and
	\begin{center}
		\label{eq:lambda}
		\begin{equation}
			\lambda(\omega) = 2\int_{0}^{\omega}\frac{\alpha^2F(\omega)}{\omega}d\omega,	
		\end{equation}
	\end{center}
	respectively, where $\gamma_{qv}$ is the phonon linewidth associated with momentum $q$ and branch index $v$, inverse of which represents the phonon lifetime and that signifies the EPC strength. $\omega_{qv}$ is the phonon frequency and $N(E_F)$ is the electron density of states (DOS) at the Fermi level. The temperature dependent superconducting gap and DOS are calculated using anisotropic Migdal-Eliashberg theory \cite{margine2013anisotropic}. The Allen-Dynes modified McMillan equation \cite{allen1975transition,mcmillan1968transition} is also employed for the calculation of $T_c$, where an effective moderate Coulomb potential $\mu_{c}^*=0.1$ is used.
	
	\begin{figure}[!htb]
		\centering
		\begin{tabular}{l}
			\includegraphics[scale=0.18]{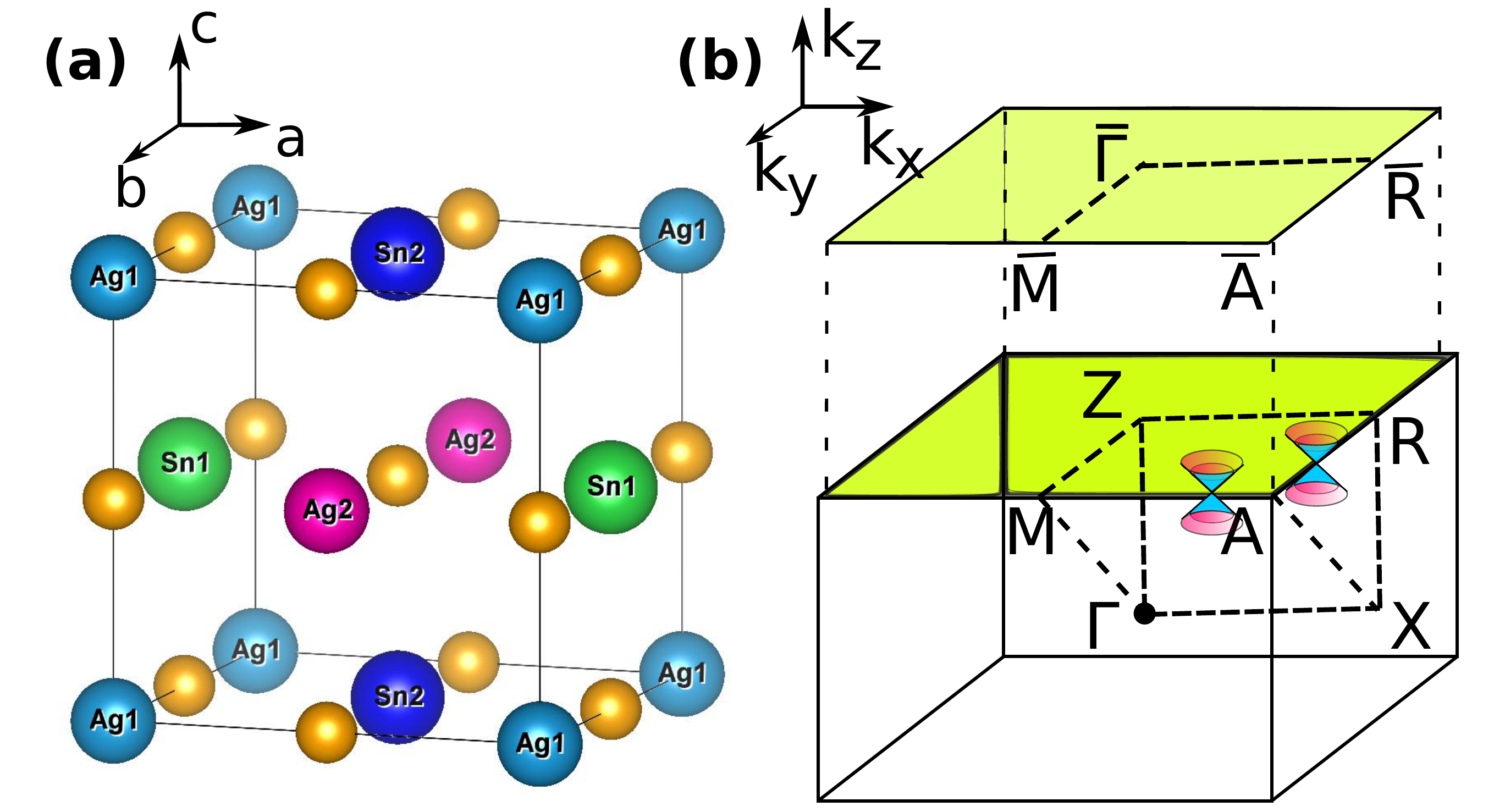}
		\end{tabular}
		\caption{(a) The crystal structures of AgSnSe$_2$, and (b) the first Brillouin zone. Two degenerate cones in (b) along $M-A$ and $R-A$ represent typical band crosiings in the bulk band structure. A 2D Brilluoin zone (BZ) surface projected on (001) plane is also shown.}
		\label{fig:struct}
	\end{figure}
	
	\section{\label{sec:elec}Crystal and electronic structures}
	
	\begin{figure*}[!htb]
		\centering
		\begin{tabular}{l}
			\hspace{-1cm}\includegraphics[scale=0.37]{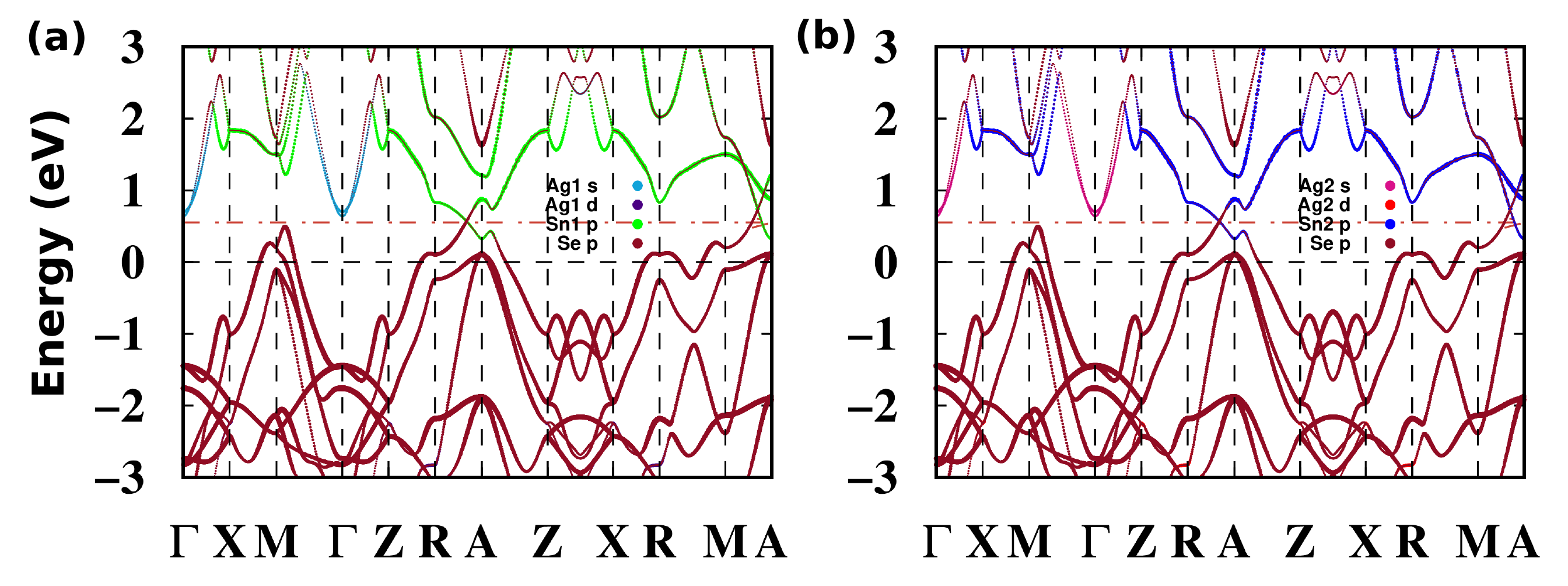}
		\end{tabular}
		\caption{The orbital projected band structures of AgSnSe$_2$ along the high-symmetry path in the BZ shown in the Fig. \ref{fig:struct}(b). (a) and (b) are orbital projected bands for atomic species at different sites shown in Fig. \ref{fig:struct}(a). The colors are labeled in the figure itself.}
		\label{fig:bands}
	\end{figure*}
	
	Though AgSnSe$_2$ show a three dimensional (3D) rocksalt NaCl and SnSe-type (space group $Fm-3m$ (225)) cubic crystal structure \cite{johnston1977superconducting,zhi2013anomalous,kataria2023superconducting} our calculated relaxed crystal structure has a slightly increased value along the third axis, with $a = b = 5.717$\AA\, and $c=5.777$\AA\,, having a tetragonal geometry with space group $P4/mmm$ (123). The inversion symmetry in this space group is preserved. Moreover, it is also protected by $C_2\rightarrow (\bar{x}(x),\bar{y}(y),z(\bar{z}))$, $C_4\rightarrow(\bar{y}(y),x(\bar{y}),z(\bar{z}))$ rotations, all three principal mirror planes $[M_{XY}\rightarrow(x,y,\bar{z});M_{XZ}\rightarrow(x,\bar{y},z); M_{YZ}\rightarrow(\bar{x},y,z)]$, and the diagonal mirror planes $[M_{D}\rightarrow(y(\bar{y}),x(\bar{x}),z(\bar{z}))]$, where the values in the inner brackets are nothing but another possibility. This tetragonal geometry is supported by an another experiment by Y. Naijo \textit{et al.} \cite{naijo2020unusual}. The crystal structure is shown in Fig. \ref{fig:struct}(a). As shown in Fig. \ref{fig:struct}(a), Ag atoms are present at the corner and at the face center while the two Sn atoms are present at the remaining two faces. Se atoms are found to be at the edge centers forming an face-centered lattice. Here, it is important to mention that experimentally Ag atoms are doped with Sn atoms at the edge centers with 1:1 ratio \cite{zhi2013anomalous}. Theoretically, the crystal structures are impossible to simulate with that configuration, and so we model our unit-cell as shown in Fig. \ref{fig:struct}(a) which is stoichiometrically similar with the experimental crystal geometry. We labeled Ag atom at the corner as Ag1 while the one at the two face centers as Ag2. Similarly, Sn atoms are labeled with Sn1 and Sn2 present at the other four faces though they are identical. The Brillouin zone (BZ) of $Ag-$doped SnSe is quite different from pristine SnSe, albeit their lattice geometries being similar. Sn$X$, where, $X=$ S, Se, Te, have a face-centered-cubic BZ \cite{sun2013rocksalt} while BZ of AgSnSe$_2$ has a cuboid shape as shown in Fig. \ref{fig:struct}(b).
	
	We calculate the band structure of AgSnSe$_2$ compound and show in the Fig. \ref{fig:bands}. There are six valence bands (three degenerate bands with a degeneracy of two) which crosss the Fermi level (FL), indicating that the system is metallic. Out of six, the lowest two bands form two small hole pockets around the BZ corner, $A$. The next two degenerate bands move along the edges of the BZ, $R-A$ and $M-A$, while the top two bands cross the FL along $X-M-\Gamma$, $Z-R-A-Z$ and $X-R-M-A$ directions. Along the latter directions on the (001) surface, the top two valence bands form two degenerate Dirac cones with the lowest two conduction bands. The orbital projections are illustrated for different atomic species at different sites in the unit-cell, in separate figures. The band structure clearly shows that the valence band region is completely occupied by the Se $p-$orbitals while the conduction bands populated by the Ag $s$ and Sn $p-$orbitals. The first two conduction bands corresponds to the two Ag atoms, and they are degenerate along the surfaces of the BZ, \textit{e.g.}, $Z-R-A-Z$ and $X-R-M-A$, while the degeneracy is lifted inside the BZ, \textit{e.g.}, $M-\Gamma-Z$ and $Z-X$. The degeneracy in the former cases is due to the symmetries involved in the system, namely the $C_4$ rotations and the mirror planes. This shows that even Ag1 and Ag2 are present at the different atomic positions in the unit cell they exhibit identical electronic states on the surfaces of the BZ. The same is true for Sn1 and Sn2. The orbital contributions of Ag $s-$orbitals along $\Gamma-X$, $\Gamma-M$ and $\Gamma-Z$ path also signifies that the these orbitals are contributing only along lines which cross the centre of the Brillouin zone (BZ) (See Fig. \ref{fig:struct}(b)). Rest of the BZ is occupied by the $p-$orbitals of Sn and Se. We will see the effect of SOC in the next section.
	
	Speaking of the skipping of +1 valence of Sn, it is evident from the orbital projections in the band structure that both Sn1 and Sn2 contribute equally in each band in the conduction band region, and it clarifies they are identical. The lowest two conduction bands belong to Sn1 and Sn2 and they are degenerate along the Brillouin zone boundaries, $Z-R-A-Z$ and $X-R-M-A$ as shown by by the green and blue projections in Fig. \ref{fig:bands}(a) and Fig. \ref{fig:bands}(b), respectively. Next we perform the bader charge analysis \cite{henkelman2006fast} for the compund and observe no charge disproportionation of Sn1 and Sn2 which verifies no charge difference between the two (valence  $= +0.906\,e$). The absence of disproportionation is also reported using Sn-NMR spectra by Y. Naijo \textit{et al.} \cite{naijo2020unusual}.
	
	\begin{figure*}[!htb]
		\centering
		\begin{tabular}{l}
			\includegraphics[scale=1]{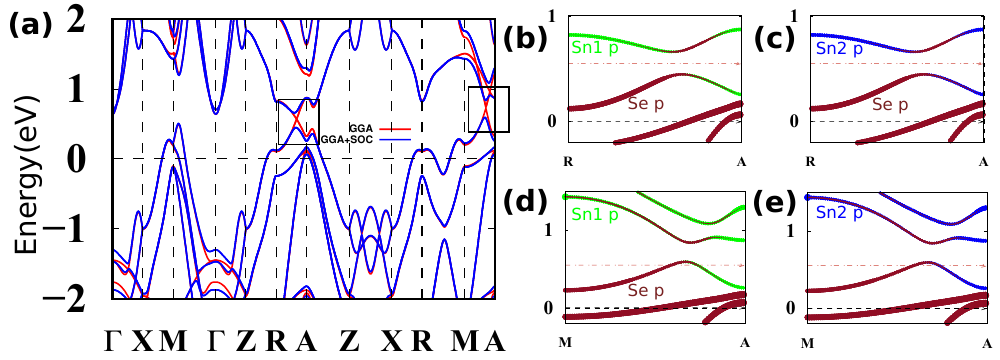}
		\end{tabular}
		\caption{(a) The band structure without (red) and with (blue) SOC. The avoided band crossings with SOC is depicted along $R-A$ and $M-A$ high symmetry directions inside the black boxes. The band inversions along (b,c) $R-A$ and (d,e) $M-A$ are shown for (b,d) Sn1 and (c,e) Sn2 in an enlarge view.}
		\label{fig:bandssoc}
	\end{figure*}
	\section{\label{sec:topo}Topological properties}
	To the best of our knowledge, the topological behavior of AgSnSe$_2$ is yet to be explored. Here, we present the topological nature of this $Ag-$doped system. As we mention in the introduction of this paper that SnSe is a relatively light weight TCI \cite{fu2011topological,sun2013rocksalt}. TCIs are characterized by (i) protected crystal symmetry ($C_4$ or $C_6$ and the mirror planes \cite{fu2011topological}), (ii) strong hybridization between $s$ and $p$ orbitals of cation and anion, (iii) large SOC strength but not a necessary condition, (iv) band inversion even without SOC (not necessarily), and more fundamentally, (v) even number of Dirac cones on the surface perpendicular to the mirror symmetric planes ($(110)$, in case of SnSe \cite{sun2013rocksalt}). In TCIs, the metallic surface states are protected by the crystal symmetry and the time-reversal symmetry ($\mathcal{TRS}$) is obsolete. Moreover, the band degeneracies in TCIs are \textit{quadratic} unlike the linearly dispersed surface states in TIs \cite{chong2008effective,sun2009topological}. We will check the said features for AgSnSe$_2$ in order to decern the topological character of the system. It is important to remark that in AgSnSe$_2$, both inversion and $\mathcal{TRS}$ are preserved. The point-group symmetries, such as $C_4$ and (110) mirror planes are also present in the system which we already discussed in Sec. \ref{sec:elec}.
	
	Before classifying the topological properties of AgSnSe$_2$ let us first revisit the electronic structure. As shown in the previously, the two valence and two conduction bands cross along $R-A$ and $M-A$ high symmetry directions, at 0.55 eV above the FL (Shown by the horizontal red line in Fig. \ref{fig:bands}). The bands form Dirac cones, which basically implies that there are band crossings in the BZ that occur only along the edges of the BZ at some non-time-reversal-invariant momenta (See Fig. \ref{fig:bands}), and thus protected by $C_4$ rotational symmetry. Now, whether these band crossings are topologically trivial or not we perform the band structure calculations incorporating the SOC. The band structure with SOC clearly shows the gap openings between the valence band maximum (VBM) and conduction band minimum (CBM) of $\sim 200$ meV along the $R-A$ and $M-A$, as shown in Fig. \ref{fig:bandssoc}. In order to check the band-inversions, we plot the orbital-projected bands with SOC and present them in Fig. \ref{fig:bandssoc}(b-e). One can clearly notice the band-inversions in the two directions via the transfer of orbital character from valance bands to the conduction bands. The band-inversions along the two high symmetry paths are shown in Fig. \ref{fig:bandssoc}(b,c) and Fig. \ref{fig:bandssoc}(d,e), respectively. The Se $p-$orbital character in the valence band extends from the $R$ point and intermixes with the conduction band towards the $A$ point. Similarly, the Sn1 and Sn2 $p-$orbital characters in conduction bands blend with the valance bands from $R$ to $A$ direction. The same trend of band-inversion is also observed for TCIs with similar band gap, mainly, rocksalt chalcogenides, Sn$X$ \cite{sun2013rocksalt} and Pb$X$ \cite{barone2013pressure}, where, $X=$ S, Se, Te. It is interesting to point out that the band inversion and transfer of orbital character are present even without incorporating SOC as shown in Fig. \ref{fig:bands} along $R-A$ and $M-A$, essentially supporting AgSnSe$_2$ to be a TCM. It is not necessary to revise that AgSnSe$_2$ is a TCM owing to the presence of electronic states at the FE and the bulk band crossings are 0.55 eV above the FL. The TCM state of the materials is not explored amply, except some oxide-perovskite iridates \cite{chen2015topological,kim2015surface}.

	\begin{figure*}[!htb]
		\centering
		\begin{tabular}{l}
			\includegraphics[scale=0.2]{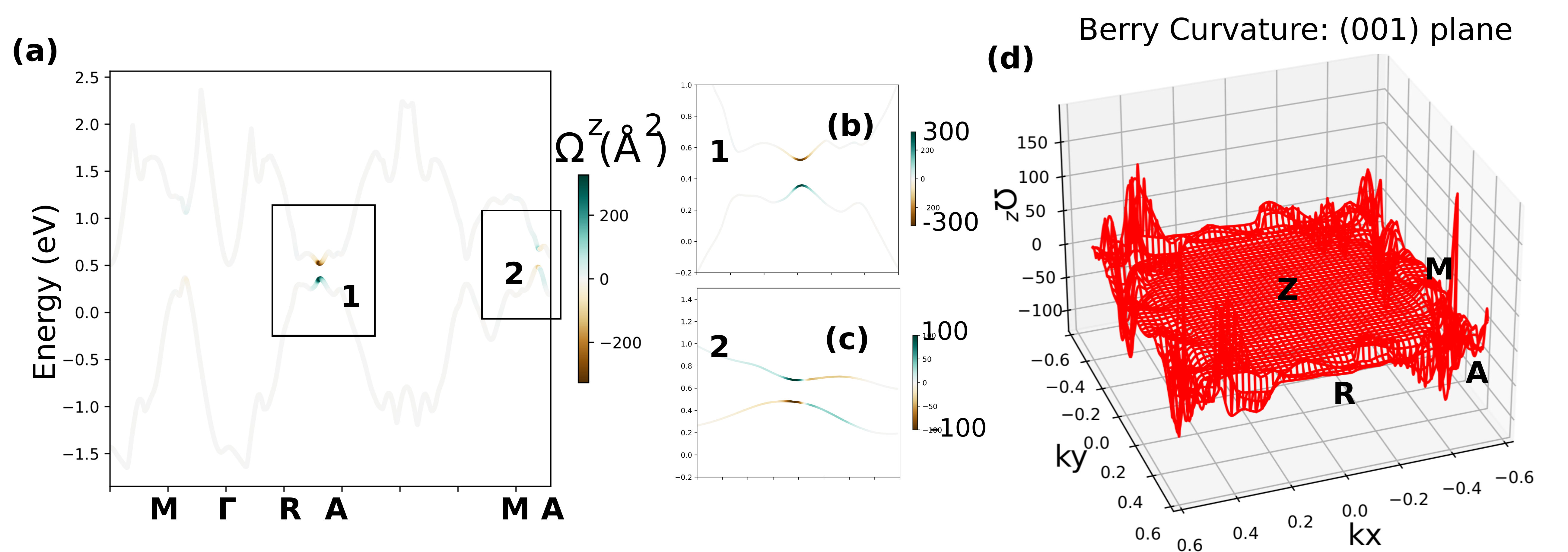}
		\end{tabular}
		\caption{(a) The Berry curvature (BC) is projected on the top valence band and lowest conduction along the high symmetry path. The regions with non-zero BC are shown in the boxes and the close views are shown in (b-d). BC in the $k_z = \pi/a$ plane is shown in (e).}
		\label{fig:berry}
	\end{figure*}
	
	In order to determine the topological character of these avoided band crossings we calculate the Berry curvature (BC) using relation \cite{berry1984quantal,xiao2010berry},
	\begin{center}
		\begin{equation}
			\label{eq:berry}
			\Omega^z  = -2 Im \sum_{n\neq v} \frac{<n_{\mathbf{k}}|\nabla_{k_x} H(\mathbf{k})|v_{\mathbf{k}}> <v_{\mathbf{k}}|\nabla_{k_y} H(\mathbf{k})|n_{\mathbf{k}}>}{(\epsilon_{n\mathbf{k}} - \epsilon_{v\mathbf{k}})^2}
		\end{equation}
	\end{center}
	where, $|n_{\mathbf{k}}>$ and $|v_{\mathbf{k}}>$ are the Bloch wavefunctions for $n^{th}$ and $v^{th}$ band. $H(\vec{k})$ is the $\mathbf{k}-$space Hamiltonian matrix which is calculated numerically using the Fourier transformation of real space Hamiltonian obtained from the MLWFs. The sum is taken over all the occupied bands below the red dashed line in Fig. \ref{fig:bands}. We plot the BC along the high symmetry path and also in the $k_z = \pi/a$ plane. We choose this particular plane for the calculation of BC as the avoided band crossing occurs only along the edges of the BZ. $R-A$ and $M-A$, as shown in Fig. \ref{fig:struct}(b), are at the edges of the BZ. We choose $k_z = \pi/a$ surface of the BZ to evaluate BC. The BC calculated along the high symmetry path shows that the magnitude is non-zero wherever the avoided band crossing occurs (Fig. \ref{fig:berry}(a-c)). The highest magnitude of BC is along $R-A$ and it is opposite for valence and conduction band, which is shown in Fig. \ref{fig:berry}(b) indicated by region-1. The BC in region-2 has comparatively small magnitude. Moreover, the BC calculated in the $k_z = \pi/a$ plane is shown in Fig. \ref{fig:berry}(d). It can be seen that the edges of the BZ show non-zero BC.
	
	\begin{figure}[!htb]
		\centering
		\begin{tabular}{l}
			\includegraphics[scale=0.22]{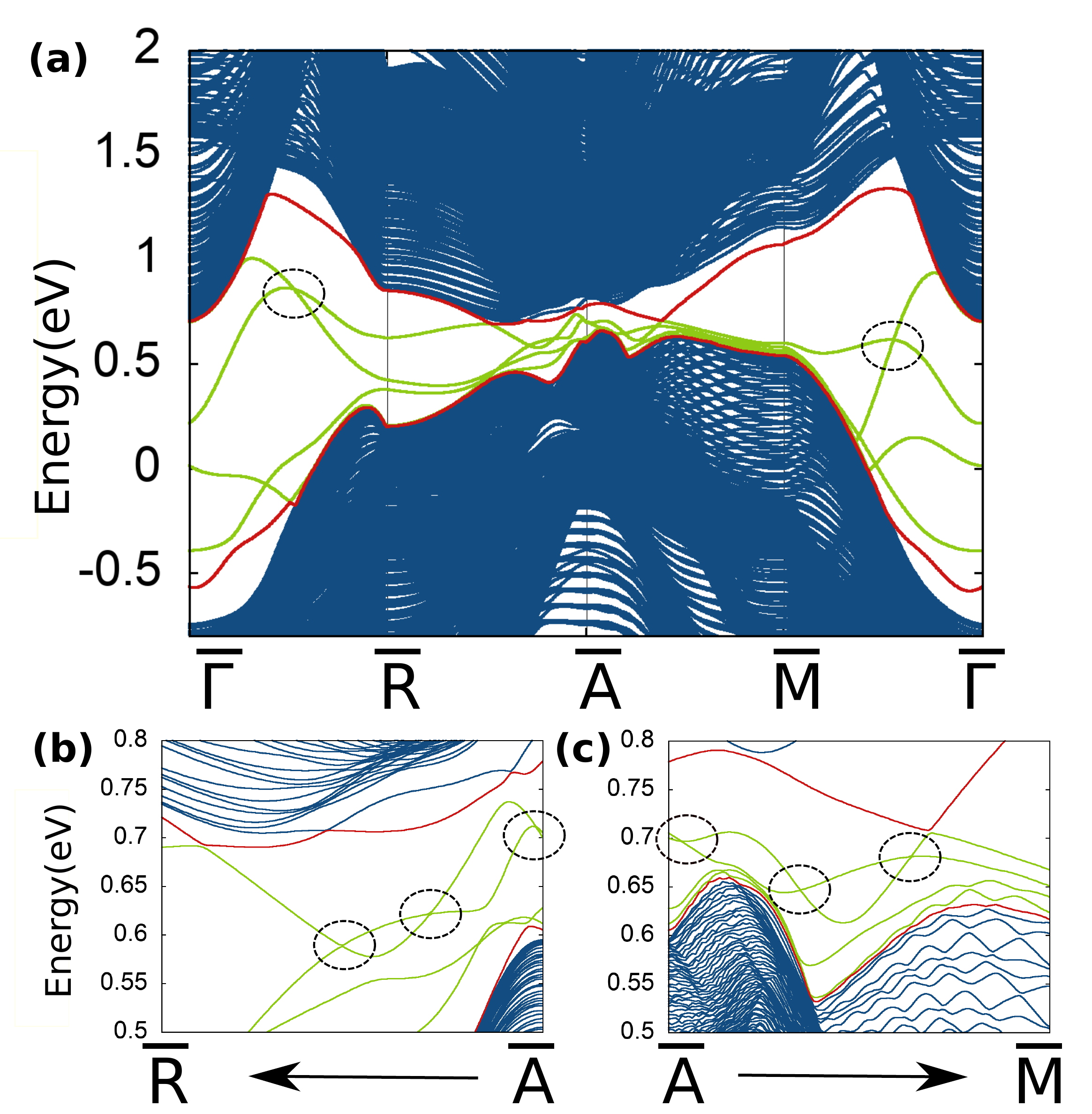}
		\end{tabular}
		\caption{(a) The surface states on the (001) projected surface of the BZ. Dirac crossings are shown by the black circles.}
		\label{fig:sstates}
	\end{figure}
	
	\begin{figure*}[!htb]
		\centering
		\begin{tabular}{l}
			\includegraphics[scale=0.8]{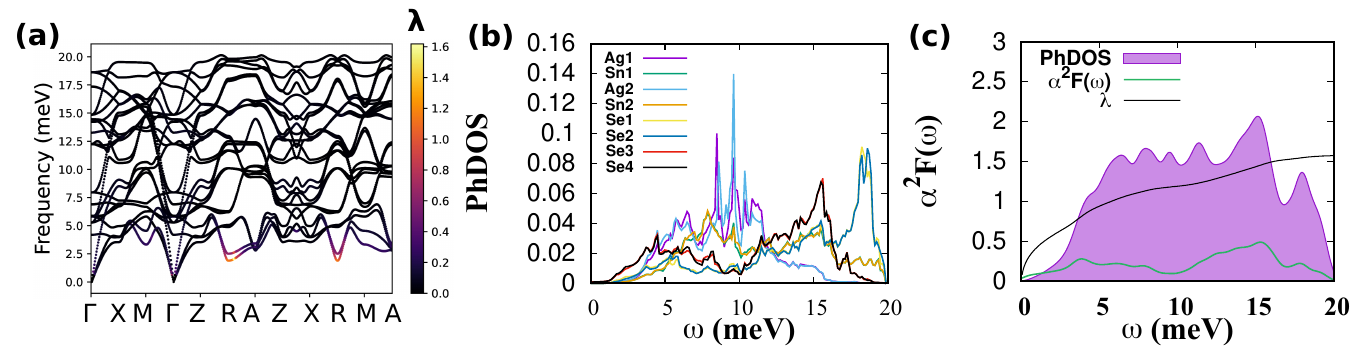}
		\end{tabular}
		\caption{(a) The EPC resolved phonon spectrum, (b) spectral function $\alpha^2F(\omega)$, total phonon DOS and cummulative EPC strength, and (c) Partial atom-resolved phonon DOS are shown.}
		\label{fig:phdos}
	\end{figure*}
	
	A non-zero Chern number ($\mathcal{C}$) indicates a non-trivial topological nature in the system \cite{haldane1988model}. Having non-zero BC for our system we calculate the $\mathcal{C}$ in the same plane in which we evaluated BC using relation, $\mathcal{C} = \frac{1}{2\pi}\int_{BZ}^{}d\mathbf{k}\Omega^z(\mathbf{k})$,
	where, the integration is taken over the $k_z = \pi/a$ plane of the BZ. The calculated value of $\mathcal{C} = 1$. Next, we calculate $\mathbb{Z}_2$ invariant. We find that for (001) plane the $\mathbb{Z}_2 = 1$ showing a non-trivial topology with $\mathbb{Z}_2$ index $(v_0;v_1,v_2,v_3) = (1;1,1,0)$. The non-zero $\mathbb{Z}_2$ indicates symmetry protected edge states in TCI \cite{fu2011topological}. It is important to mention that the TCIs are characterized by \textit{mirror} Chern number ($n_{\mathcal{M}}$) \cite{teo2008surface,fu2011topological,junwei2013two,alexandradinata2016berry}. A non-zero $n_{\mathcal{M}}$ refers to some non-trivial topology and mirror symmetry-protected surface states in TCIs. Therefore, we use the tight-binding Hamiltonian obtained from {\footnotesize WANNIER90}, and prepare a slab of 40 unit cells along $c-$axis. We the perform slab calculations to calculate the surface states using MLWFs as implemented in {\footnotesize WannierTools} \cite{wu2017wanniertools}. We indeed observe the edge states in the spectrum of the slab geometry as shown in Fig. \ref{fig:sstates}. The surface states show Dirac cones along the edges, away from the time-reversal-invariant momenta ({\small TRIM}), and around the positions of bulk band crossings, i.e., along $R-A$ and $M-A$ (respective {\small TRIM} are $\bar{R}$, $\bar{A}$ and $\bar{M}$ on the surface BZ). These edge modes are also known as helical hinge modes as disseminated for the higher order topological insulators (HOTIs) \cite{schindler2018higher}. One can observe multiple Dirac cones in the spectrum of surface states. These Dirac cones are shown in black circles in Fig. \ref{fig:sstates}(a) (along $\bar{A}-\bar{R}$) and Fig. \ref{fig:sstates}(a) (along $\bar{A}-\bar{M}$). In addition, we also observe band crossings along the two mirror planes, $M_X$ and $M_Y$, namely, $\Gamma-R$ and $\Gamma-M$, respectively (not shown), which further support towards the TCM nature of AgSnSe$_2$. $M_X$ and $M_Y$ planes pass through the center of the BZ and parallel to $k_z$. These Dirac cones along the mirror planes on (001) surface have been observed previously for rocksalt Sn$X$ TCIs \cite{junwei2013two,wang2013nontrivial} with $n_{\mathcal{M}}=-2$. Interestingly, the bulk band structure does not show any avoided band crossing along this direction, $Z-R(M)$ (See Fig. \ref{fig:bandssoc}(a)). Conversely, there are conducting surface states in the slab geometry, protected by the mirror planes, $M_X$ and $M_Y$. The existence of two pairs of counter-propagating, spin-polarized surface states with opposite mirror eigenvalues (opposite chirality) along $\bar{R}(\bar{M})\leftarrow\bar{\Gamma}\rightarrow-\bar{R}(-\bar{M})$ dictates $n_{\mathcal{M}}=-2$ for the TCM (and TCIs). Moreover, in the presence of pressure (reduced lattice constant), these symmetry protected surface states along $\bar{\Gamma}--\bar{R}(-\bar{M})$ disappear, and a gap opens up, which further supports a TCM trait \cite{sun2013rocksalt}. The 3D TCIs of Sn$X$ class have been investigated for quantum spin Hall effect (QSHE) theoretically \cite{liu2015electrically,safaei2015quantum}. The surface states in AgSnSe$_2$ can engender quantum spin Hall conductivity in the system, which, to the best of our knowledge, remains unexplored. More experimental evidence is required to substantiate QSHE.

	\section{\label{sec:sc}Anisotropic superconducting properties}
	SnSe has been investigated thoroughly for its topological properties. The physical properties of this material tuned via pressure is known in the literature \cite{hsieh2012topological,sun2013rocksalt}. Doping is another handle to manipulate the intrinsic properties of the system. Hole doping is used for a transition toward a superconducting state, and AgSnSe$_2$ has been experimentally investigated for its superconductivity (SC) \cite{zhi2013anomalous,naijo2020unusual,kataria2023superconducting,matsuura2022valence}. Using Migdal-Eliashberg Theory (MET) we perform electron-phonon calculations followed by anisotropic superconductivity calculations. In Fig. \ref{fig:phdos}(a), the phonon spectra is shown. The phonon softening along $Z-R-A$ and $X-R-M$ in the acoustic mode is mainly responsible for the EPC in the system, the strength of which is shown by the color scale. This particular phonon branch belongs to the Se3/Se4 atoms as presented in the partial phonon DOS in Fig. \ref{fig:phdos}(b) by the red/black curve. We further calculate the spectral function $\alpha^2F(\omega)$ and show in Fig. \ref{fig:phdos}(c) along with the integrated EPC, $\lambda$. The EPC strength shown by the black curve. It gradually increases in the whole frequency region and saturates at $1.65$ which is also the maximum strength in the color scale of Fig. \ref{fig:phdos}(a). There is no sudden change in $\lambda$ which signifies that the system AgSnSe$_2$ is a one-gap superconductor which we will show next. The $\alpha^2F(\omega)$ and the phonon DOS qualitatively coincide with each others in the whole frequency range. Interestingly, if we compare $\alpha^2F(\omega)$ with the partial phonon DOS of Se3 (red) and Se4 (black) they follow the same pattern in the entire frequecy range. This also corroborate our previous statement that the Se3 and Se4 modes are responsible for the EPC in the system. Further investigations of crystal orbital Hamiltonian population ({\footnotesize COHP}) can determine the bonding strength in the system.

	\begin{figure}[!htb]
		\centering
		\begin{tabular}{l}
			\includegraphics[scale=0.6]{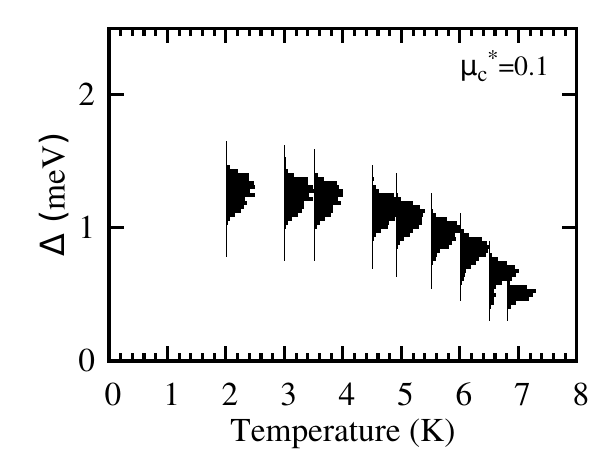}
		\end{tabular}
		\caption{The energy distribution of the gap, $\Delta$, as a function of temperature at $\mu_{c}^* = 0.1$. One-gap SC is obvious from the plot.}
		\label{fig:gap}
	\end{figure}
	
	Next we calculate the superconducting gap function using MET. We first evaluate the critical temperature, $T_c$ using modified McMillan's formula called Allen-Dynes $T_c$ \cite{mcmillan1968transition,allen1975transition},
	
	\begin{equation}
		T_c  = \frac{\omega_{log}}{1.2}\, exp\left[\frac{-1.04(1+\lambda)}{\lambda - \mu^*_c(1+0.62\lambda)}\right].
	\end{equation} 
	Here, $\omega_{log}$ is the logarithmic average of frequency with the unit of temperature, $\lambda$ is a dimensionless parameter called EPC strength, and $\mu^*_c$ is the effective Coulomb repulsion parameter. The $T_c$ calculated using this relation is $3.1$K at $\mu^*_c = 0.1$ and $\lambda=1.65$, which is very close to the experimentally found $T_c = 4.91$K. The anisotropic MET, on the other hand, gives $T_c = 7.2$K as shown in the superconducting gap function plotted in Fig \ref{fig:gap}. Though the magnitude of $T_c$ is slightly overestimated using MET it is comparable to the experimental $T_c$ indicating negligible anisotropy in AgSnSe$_2$. Apparently, the EPC strength and $T_c$ can be tuned by varying $\mu^*_{c}$, and SOC can further modify the zero temperature gap, $\Delta(0)$ and $T_c$, as reported by us previously \cite{patel2024electron}. This close estimate of $T_c$ indicates that EPC is playing the major role for the superconductivity in the system. Additionally, MET predicts that AgSnSe$_2$ is a one-gap superconductor unlike what is reported by one of the experiments \cite{kataria2023superconducting} claiming it to be a valence-skipping two-gap superconductor. Further analysis suggests that the BCS ratio, $\Delta(0)/k_BT_C=2.155$, which corresponds to the higher gap reported in the experiment \cite{kataria2023superconducting}. This value is higher than the standard BCS ratio for the weak coupling ($1.76$). This larger value again underpins a strong EPC in AgSnSe$_2$.
	
	Morever, valence-skipping (VS) is a unique phenomenon especially ocurring in $5s$ and $6s$ states. While some exotic properties such as enhanced SC and charge Kondo effect can be explained via the resulting attractive-$U$ \cite{dzero2005superconductivity}, but this is not a necessary condition. While SC is a common feature in VS materials there are other properties which might open a new avenue for material applications. Additionally, the Fermi energy (FE) can be tuned by $Ag$-doping concentration and consequently the physical properties of AgSnSe$_2$ as well. As mentioned earlier, $Ag$ shifts the chemical potential (FE) down into the valence band region. Some other element can be used as electron doping, to bring back the chemical potential in the middle of the band gap and tune the topological properties. Gate voltage is an another choice for the experimentalists. AgSnSe$_2$ can be studied for its thermoelectric performance. 
	
	\section{\label{sec:conclusions}Conclusions}
	In conclusion, we investigated the electronic properties of AgSnSe$_2$ compound using first-principles density functional theory. The occupied bands are dominated by Se $p$ orbitals while there are different contributions in the conduction band region. The lowest conduction band is predominantly from Sn $p$ orbitals in the whole Brillouin zone except around the $\Gamma$ point where it is contributed by the Ag $s$ orbitals. We divide the atomic species in two types and interestingly both atomic species have equal contributions of orbitals in the conduction band region. After identifying the band-inversion character in the band structure around $0.55$ eV above the Fermi level along $R-A$ and $M-A$ $k$-path, which forms the edges of the BZ, we calculated the Berry curvature on one face (top) of the BZ. We find non-zero Berry curvature, which is mainly concentrated along the edges of (001) plane. Further analyses of symmetry protected surface states and non-zero \textit{mirror} Chern number, $n_\mathcal{M}$, gaurantees that AgSnSe$_2$ is a TCM. Next, because of the hole doping by Ag, the valence bands give rise to Fermi surface sheets at the FL. We, therefore, calculate the superconducting properties of AgSnSe$_2$ using Migdal-Eliashberg theory. We find that the main contribution in the spectral function $\alpha^2F(\omega)$, is coming from the acoustic phonon modes of Se3 and Se4, which soften around the $R$ point. Our calculated superconducting gap function using MET suggests single-gap SC in AgSnSe$_2$ in contrast to one of the recent experimental study, and a vital role of EPC. A large number of valleys in valence and conduction bands suggests AgSnSe$_2$ is a good candidate to explore the thermoelectric properties and possible exciton dynamics.
	
	\section{\label{sec:ackn}Acknowledgements}
	The authors acknowledge National Supercomputing Mission (NSM) for providing computing resources of 'PARAM Shakti' at IIT Kharagpur, which is implemented by C-DAC and supported by the Ministry of Electronics and Information Technology (MeitY) and Department of Science and Technology (DST), Government of India. AT acknowledges useful discussions on the experimental evidence in favor of possible valence-skipping in AgSnSe$_2$ with Ravi Prakash Singh.

	\bibliography{refs}

\begin{thebibliography}{51}%
\makeatletter
\providecommand \@ifxundefined [1]{%
 \@ifx{#1\undefined}
}%
\providecommand \@ifnum [1]{%
 \ifnum #1\expandafter \@firstoftwo
 \else \expandafter \@secondoftwo
 \fi
}%
\providecommand \@ifx [1]{%
 \ifx #1\expandafter \@firstoftwo
 \else \expandafter \@secondoftwo
 \fi
}%
\providecommand \natexlab [1]{#1}%
\providecommand \enquote  [1]{``#1''}%
\providecommand \bibnamefont  [1]{#1}%
\providecommand \bibfnamefont [1]{#1}%
\providecommand \citenamefont [1]{#1}%
\providecommand \href@noop [0]{\@secondoftwo}%
\providecommand \href [0]{\begingroup \@sanitize@url \@href}%
\providecommand \@href[1]{\@@startlink{#1}\@@href}%
\providecommand \@@href[1]{\endgroup#1\@@endlink}%
\providecommand \@sanitize@url [0]{\catcode `\\12\catcode `\$12\catcode
  `\&12\catcode `\#12\catcode `\^12\catcode `\_12\catcode `\%12\relax}%
\providecommand \@@startlink[1]{}%
\providecommand \@@endlink[0]{}%
\providecommand \url  [0]{\begingroup\@sanitize@url \@url }%
\providecommand \@url [1]{\endgroup\@href {#1}{\urlprefix }}%
\providecommand \urlprefix  [0]{URL }%
\providecommand \Eprint [0]{\href }%
\providecommand \doibase [0]{https://doi.org/}%
\providecommand \selectlanguage [0]{\@gobble}%
\providecommand \bibinfo  [0]{\@secondoftwo}%
\providecommand \bibfield  [0]{\@secondoftwo}%
\providecommand \translation [1]{[#1]}%
\providecommand \BibitemOpen [0]{}%
\providecommand \bibitemStop [0]{}%
\providecommand \bibitemNoStop [0]{.\EOS\space}%
\providecommand \EOS [0]{\spacefactor3000\relax}%
\providecommand \BibitemShut  [1]{\csname bibitem#1\endcsname}%
\let\auto@bib@innerbib\@empty
%</preamble>
\bibitem [{\citenamefont {Taraphder}\ and\ \citenamefont
  {Coleman}(1991)}]{taraphder1991heavy}%
  \BibitemOpen
  \bibfield  {author} {\bibinfo {author} {\bibfnamefont {A.}~\bibnamefont
  {Taraphder}}\ and\ \bibinfo {author} {\bibfnamefont {P.}~\bibnamefont
  {Coleman}},\ }\bibfield  {title} {\bibinfo {title} {{Heavy-fermion behavior
  in a negative-U Anderson model}},\ }\href
  {https://doi.org/10.1103/PhysRevLett.66.2814} {\bibfield  {journal} {\bibinfo
   {journal} {Phys. Rev. Lett.}\ }\textbf {\bibinfo {volume} {66}},\ \bibinfo
  {pages} {2814} (\bibinfo {year} {1991})}\BibitemShut {NoStop}%
\bibitem [{\citenamefont {Taraphder}\ \emph {et~al.}(1995)\citenamefont
  {Taraphder}, \citenamefont {Krishnamurthy}, \citenamefont {Pandit},\ and\
  \citenamefont {Ramakrishnan}}]{taraphder1995negative}%
  \BibitemOpen
  \bibfield  {author} {\bibinfo {author} {\bibfnamefont {A.}~\bibnamefont
  {Taraphder}}, \bibinfo {author} {\bibfnamefont {H.~R.}\ \bibnamefont
  {Krishnamurthy}}, \bibinfo {author} {\bibfnamefont {R.}~\bibnamefont
  {Pandit}},\ and\ \bibinfo {author} {\bibfnamefont {T.~V.}\ \bibnamefont
  {Ramakrishnan}},\ }\bibfield  {title} {\bibinfo {title} {{Negative-U extended
  Hubbard model for doped barium bismuthates}},\ }\href
  {https://doi.org/10.1103/PhysRevB.52.1368} {\bibfield  {journal} {\bibinfo
  {journal} {Phys. Rev. B}\ }\textbf {\bibinfo {volume} {52}},\ \bibinfo
  {pages} {1368} (\bibinfo {year} {1995})}\BibitemShut {NoStop}%
\bibitem [{\citenamefont {Taraphder}\ \emph {et~al.}(1996)\citenamefont
  {Taraphder}, \citenamefont {Pandit}, \citenamefont {Krishnamurthy},\ and\
  \citenamefont {Ramakrishnan}}]{taraphder1996exotic}%
  \BibitemOpen
  \bibfield  {author} {\bibinfo {author} {\bibfnamefont {A.}~\bibnamefont
  {Taraphder}}, \bibinfo {author} {\bibfnamefont {R.}~\bibnamefont {Pandit}},
  \bibinfo {author} {\bibfnamefont {H.~R.}\ \bibnamefont {Krishnamurthy}},\
  and\ \bibinfo {author} {\bibfnamefont {T.~V.}\ \bibnamefont {Ramakrishnan}},\
  }\bibfield  {title} {\bibinfo {title} {The exotic barium bismuthates},\
  }\href {https://doi.org/10.1142/S0217979296000362} {\bibfield  {journal}
  {\bibinfo  {journal} {International Journal of Modern Physics B}\ }\textbf
  {\bibinfo {volume} {10}},\ \bibinfo {pages} {863} (\bibinfo {year}
  {1996})}\BibitemShut {NoStop}%
\bibitem [{\citenamefont {Varma}(1988)}]{varma1988missing}%
  \BibitemOpen
  \bibfield  {author} {\bibinfo {author} {\bibfnamefont {C.~M.}\ \bibnamefont
  {Varma}},\ }\bibfield  {title} {\bibinfo {title} {Missing valence states,
  diamagnetic insulators, and superconductors},\ }\href
  {https://doi.org/10.1103/PhysRevLett.61.2713} {\bibfield  {journal} {\bibinfo
   {journal} {Phys. Rev. Lett.}\ }\textbf {\bibinfo {volume} {61}},\ \bibinfo
  {pages} {2713} (\bibinfo {year} {1988})}\BibitemShut {NoStop}%
\bibitem [{\citenamefont {Johnston}\ and\ \citenamefont
  {Adrian}(1977)}]{johnston1977superconducting}%
  \BibitemOpen
  \bibfield  {author} {\bibinfo {author} {\bibfnamefont {D.}~\bibnamefont
  {Johnston}}\ and\ \bibinfo {author} {\bibfnamefont {H.}~\bibnamefont
  {Adrian}},\ }\bibfield  {title} {\bibinfo {title} {{Superconducting and
  normal state properties of Ag$_{1- x}$Sn$_{1+ x}$Se$_{2- y}$}},\ }\href
  {https://doi.org/10.1016/0022-3697(77)90080-4} {\bibfield  {journal}
  {\bibinfo  {journal} {Journal of Physics and Chemistry of Solids}\ }\textbf
  {\bibinfo {volume} {38}},\ \bibinfo {pages} {355} (\bibinfo {year}
  {1977})}\BibitemShut {NoStop}%
\bibitem [{\citenamefont {Nasredinov}\ \emph {et~al.}(1999)\citenamefont
  {Nasredinov}, \citenamefont {Nemov}, \citenamefont {Masterov},\ and\
  \citenamefont {Seregin}}]{nasredinov1999mossbauer}%
  \BibitemOpen
  \bibfield  {author} {\bibinfo {author} {\bibfnamefont {F.}~\bibnamefont
  {Nasredinov}}, \bibinfo {author} {\bibfnamefont {S.}~\bibnamefont {Nemov}},
  \bibinfo {author} {\bibfnamefont {V.}~\bibnamefont {Masterov}},\ and\
  \bibinfo {author} {\bibfnamefont {P.}~\bibnamefont {Seregin}},\ }\bibfield
  {title} {\bibinfo {title} {M{\"o}ssbauer studies of negative-u tin centers in
  lead chalcogenides},\ }\href {https://doi.org/10.1134/1.1131091} {\bibfield
  {journal} {\bibinfo  {journal} {Fiz. Tverd. Tela}\ }\textbf {\bibinfo
  {volume} {41}},\ \bibinfo {pages} {1897} (\bibinfo {year}
  {1999})}\BibitemShut {NoStop}%
\bibitem [{\citenamefont {Kataria}\ \emph {et~al.}(2023)\citenamefont
  {Kataria}, \citenamefont {Arushi}, \citenamefont {Sharma}, \citenamefont
  {Agarwal}, \citenamefont {Pula}, \citenamefont {Beare}, \citenamefont {Yoon},
  \citenamefont {Cai}, \citenamefont {Kojima}, \citenamefont {Luke},\ and\
  \citenamefont {Singh}}]{kataria2023superconducting}%
  \BibitemOpen
  \bibfield  {author} {\bibinfo {author} {\bibfnamefont {A.}~\bibnamefont
  {Kataria}}, \bibinfo {author} {\bibnamefont {Arushi}}, \bibinfo {author}
  {\bibfnamefont {S.}~\bibnamefont {Sharma}}, \bibinfo {author} {\bibfnamefont
  {T.}~\bibnamefont {Agarwal}}, \bibinfo {author} {\bibfnamefont
  {M.}~\bibnamefont {Pula}}, \bibinfo {author} {\bibfnamefont {J.}~\bibnamefont
  {Beare}}, \bibinfo {author} {\bibfnamefont {S.}~\bibnamefont {Yoon}},
  \bibinfo {author} {\bibfnamefont {Y.}~\bibnamefont {Cai}}, \bibinfo {author}
  {\bibfnamefont {K.~M.}\ \bibnamefont {Kojima}}, \bibinfo {author}
  {\bibfnamefont {G.~M.}\ \bibnamefont {Luke}},\ and\ \bibinfo {author}
  {\bibfnamefont {R.~P.}\ \bibnamefont {Singh}},\ }\bibfield  {title} {\bibinfo
  {title} {{Superconducting ground state study of the valence-skipped compound
  ${\mathrm{AgSnSe}}_{2}$}},\ }\href
  {https://doi.org/10.1103/PhysRevB.107.174517} {\bibfield  {journal} {\bibinfo
   {journal} {Phys. Rev. B}\ }\textbf {\bibinfo {volume} {107}},\ \bibinfo
  {pages} {174517} (\bibinfo {year} {2023})}\BibitemShut {NoStop}%
\bibitem [{\citenamefont {Ren}\ \emph {et~al.}(2013)\citenamefont {Ren},
  \citenamefont {Kriener}, \citenamefont {Taskin}, \citenamefont {Sasaki},
  \citenamefont {Segawa},\ and\ \citenamefont {Ando}}]{zhi2013anomalous}%
  \BibitemOpen
  \bibfield  {author} {\bibinfo {author} {\bibfnamefont {Z.}~\bibnamefont
  {Ren}}, \bibinfo {author} {\bibfnamefont {M.}~\bibnamefont {Kriener}},
  \bibinfo {author} {\bibfnamefont {A.~A.}\ \bibnamefont {Taskin}}, \bibinfo
  {author} {\bibfnamefont {S.}~\bibnamefont {Sasaki}}, \bibinfo {author}
  {\bibfnamefont {K.}~\bibnamefont {Segawa}},\ and\ \bibinfo {author}
  {\bibfnamefont {Y.}~\bibnamefont {Ando}},\ }\bibfield  {title} {\bibinfo
  {title} {{Anomalous metallic state above the upper critical field of the
  conventional three-dimensional superconductor AgSnSe$_{2}$ with strong
  intrinsic disorder}},\ }\href {https://doi.org/10.1103/PhysRevB.87.064512}
  {\bibfield  {journal} {\bibinfo  {journal} {Phys. Rev. B}\ }\textbf {\bibinfo
  {volume} {87}},\ \bibinfo {pages} {064512} (\bibinfo {year}
  {2013})}\BibitemShut {NoStop}%
\bibitem [{\citenamefont {Naijo}\ \emph {et~al.}(2020)\citenamefont {Naijo},
  \citenamefont {Hada}, \citenamefont {Furukawa}, \citenamefont {Itou},
  \citenamefont {Ueno}, \citenamefont {Kobayashi}, \citenamefont {Mazin},
  \citenamefont {Jeschke},\ and\ \citenamefont {Akimitsu}}]{naijo2020unusual}%
  \BibitemOpen
  \bibfield  {author} {\bibinfo {author} {\bibfnamefont {Y.}~\bibnamefont
  {Naijo}}, \bibinfo {author} {\bibfnamefont {K.}~\bibnamefont {Hada}},
  \bibinfo {author} {\bibfnamefont {T.}~\bibnamefont {Furukawa}}, \bibinfo
  {author} {\bibfnamefont {T.}~\bibnamefont {Itou}}, \bibinfo {author}
  {\bibfnamefont {T.}~\bibnamefont {Ueno}}, \bibinfo {author} {\bibfnamefont
  {K.}~\bibnamefont {Kobayashi}}, \bibinfo {author} {\bibfnamefont {I.~I.}\
  \bibnamefont {Mazin}}, \bibinfo {author} {\bibfnamefont {H.~O.}\ \bibnamefont
  {Jeschke}},\ and\ \bibinfo {author} {\bibfnamefont {J.}~\bibnamefont
  {Akimitsu}},\ }\bibfield  {title} {\bibinfo {title} {Unusual electronic state
  of sn in ${\mathrm{agsnse}}_{2}$},\ }\href
  {https://doi.org/10.1103/PhysRevB.101.075134} {\bibfield  {journal} {\bibinfo
   {journal} {Phys. Rev. B}\ }\textbf {\bibinfo {volume} {101}},\ \bibinfo
  {pages} {075134} (\bibinfo {year} {2020})}\BibitemShut {NoStop}%
\bibitem [{\citenamefont {Sun}\ \emph {et~al.}(2013)\citenamefont {Sun},
  \citenamefont {Zhong}, \citenamefont {Shirakawa}, \citenamefont {Franchini},
  \citenamefont {Li}, \citenamefont {Li}, \citenamefont {Yunoki},\ and\
  \citenamefont {Chen}}]{sun2013rocksalt}%
  \BibitemOpen
  \bibfield  {author} {\bibinfo {author} {\bibfnamefont {Y.}~\bibnamefont
  {Sun}}, \bibinfo {author} {\bibfnamefont {Z.}~\bibnamefont {Zhong}}, \bibinfo
  {author} {\bibfnamefont {T.}~\bibnamefont {Shirakawa}}, \bibinfo {author}
  {\bibfnamefont {C.}~\bibnamefont {Franchini}}, \bibinfo {author}
  {\bibfnamefont {D.}~\bibnamefont {Li}}, \bibinfo {author} {\bibfnamefont
  {Y.}~\bibnamefont {Li}}, \bibinfo {author} {\bibfnamefont {S.}~\bibnamefont
  {Yunoki}},\ and\ \bibinfo {author} {\bibfnamefont {X.-Q.}\ \bibnamefont
  {Chen}},\ }\bibfield  {title} {\bibinfo {title} {{Rocksalt SnS and SnSe:
  Native topological crystalline insulators}},\ }\href
  {https://doi.org/10.1103/PhysRevB.88.235122} {\bibfield  {journal} {\bibinfo
  {journal} {Phys. Rev. B}\ }\textbf {\bibinfo {volume} {88}},\ \bibinfo
  {pages} {235122} (\bibinfo {year} {2013})}\BibitemShut {NoStop}%
\bibitem [{\citenamefont {Jin}\ \emph {et~al.}(2017)\citenamefont {Jin},
  \citenamefont {Vishwanath}, \citenamefont {Liu}, \citenamefont {Kong},
  \citenamefont {Lou}, \citenamefont {Dai}, \citenamefont {Sadowski},
  \citenamefont {Liu}, \citenamefont {Lien}, \citenamefont {Chaney},
  \citenamefont {Han}, \citenamefont {Cao}, \citenamefont {Ma}, \citenamefont
  {Qian}, \citenamefont {Wang}, \citenamefont {Dobrowolska}, \citenamefont
  {Furdyna}, \citenamefont {Muller}, \citenamefont {Pohl}, \citenamefont
  {Ding}, \citenamefont {Dadap}, \citenamefont {Xing},\ and\ \citenamefont
  {Osgood}}]{jin2017electronic}%
  \BibitemOpen
  \bibfield  {author} {\bibinfo {author} {\bibfnamefont {W.}~\bibnamefont
  {Jin}}, \bibinfo {author} {\bibfnamefont {S.}~\bibnamefont {Vishwanath}},
  \bibinfo {author} {\bibfnamefont {J.}~\bibnamefont {Liu}}, \bibinfo {author}
  {\bibfnamefont {L.}~\bibnamefont {Kong}}, \bibinfo {author} {\bibfnamefont
  {R.}~\bibnamefont {Lou}}, \bibinfo {author} {\bibfnamefont {Z.}~\bibnamefont
  {Dai}}, \bibinfo {author} {\bibfnamefont {J.~T.}\ \bibnamefont {Sadowski}},
  \bibinfo {author} {\bibfnamefont {X.}~\bibnamefont {Liu}}, \bibinfo {author}
  {\bibfnamefont {H.-H.}\ \bibnamefont {Lien}}, \bibinfo {author}
  {\bibfnamefont {A.}~\bibnamefont {Chaney}}, \bibinfo {author} {\bibfnamefont
  {Y.}~\bibnamefont {Han}}, \bibinfo {author} {\bibfnamefont {M.}~\bibnamefont
  {Cao}}, \bibinfo {author} {\bibfnamefont {J.}~\bibnamefont {Ma}}, \bibinfo
  {author} {\bibfnamefont {T.}~\bibnamefont {Qian}}, \bibinfo {author}
  {\bibfnamefont {S.}~\bibnamefont {Wang}}, \bibinfo {author} {\bibfnamefont
  {M.}~\bibnamefont {Dobrowolska}}, \bibinfo {author} {\bibfnamefont
  {J.}~\bibnamefont {Furdyna}}, \bibinfo {author} {\bibfnamefont {D.~A.}\
  \bibnamefont {Muller}}, \bibinfo {author} {\bibfnamefont {K.}~\bibnamefont
  {Pohl}}, \bibinfo {author} {\bibfnamefont {H.}~\bibnamefont {Ding}}, \bibinfo
  {author} {\bibfnamefont {J.~I.}\ \bibnamefont {Dadap}}, \bibinfo {author}
  {\bibfnamefont {H.~G.}\ \bibnamefont {Xing}},\ and\ \bibinfo {author}
  {\bibfnamefont {R.~M.}\ \bibnamefont {Osgood}},\ }\bibfield  {title}
  {\bibinfo {title} {{Electronic Structure of the Metastable Epitaxial
  Rock-Salt SnSe ${111}$ Topological Crystalline Insulator}},\ }\href
  {https://doi.org/10.1103/PhysRevX.7.041020} {\bibfield  {journal} {\bibinfo
  {journal} {Phys. Rev. X}\ }\textbf {\bibinfo {volume} {7}},\ \bibinfo {pages}
  {041020} (\bibinfo {year} {2017})}\BibitemShut {NoStop}%
\bibitem [{\citenamefont {Pletikosi{\'c}}\ \emph {et~al.}(2018)\citenamefont
  {Pletikosi{\'c}}, \citenamefont {von Rohr}, \citenamefont {Pervan},
  \citenamefont {Das}, \citenamefont {Vobornik}, \citenamefont {Cava},\ and\
  \citenamefont {Valla}}]{pletikosi2018band}%
  \BibitemOpen
  \bibfield  {author} {\bibinfo {author} {\bibfnamefont {I.}~\bibnamefont
  {Pletikosi{\'c}}}, \bibinfo {author} {\bibfnamefont {F.}~\bibnamefont {von
  Rohr}}, \bibinfo {author} {\bibfnamefont {P.}~\bibnamefont {Pervan}},
  \bibinfo {author} {\bibfnamefont {P.~K.}\ \bibnamefont {Das}}, \bibinfo
  {author} {\bibfnamefont {I.}~\bibnamefont {Vobornik}}, \bibinfo {author}
  {\bibfnamefont {R.~J.}\ \bibnamefont {Cava}},\ and\ \bibinfo {author}
  {\bibfnamefont {T.}~\bibnamefont {Valla}},\ }\bibfield  {title} {\bibinfo
  {title} {{Band Structure of the IV-VI Black Phosphorus Analog and
  Thermoelectric SnSe}},\ }\href
  {https://doi.org/10.1103/PhysRevLett.120.156403} {\bibfield  {journal}
  {\bibinfo  {journal} {Phys. Rev. Lett.}\ }\textbf {\bibinfo {volume} {120}},\
  \bibinfo {pages} {156403} (\bibinfo {year} {2018})}\BibitemShut {NoStop}%
\bibitem [{\citenamefont {Kutorasinski}\ \emph {et~al.}(2015)\citenamefont
  {Kutorasinski}, \citenamefont {Wiendlocha}, \citenamefont {Kaprzyk},\ and\
  \citenamefont {Tobola}}]{kutorasinski2015electronic}%
  \BibitemOpen
  \bibfield  {author} {\bibinfo {author} {\bibfnamefont {K.}~\bibnamefont
  {Kutorasinski}}, \bibinfo {author} {\bibfnamefont {B.}~\bibnamefont
  {Wiendlocha}}, \bibinfo {author} {\bibfnamefont {S.}~\bibnamefont
  {Kaprzyk}},\ and\ \bibinfo {author} {\bibfnamefont {J.}~\bibnamefont
  {Tobola}},\ }\bibfield  {title} {\bibinfo {title} {Electronic structure and
  thermoelectric properties of $n$- and $p$-type snse from first-principles
  calculations},\ }\href {https://doi.org/10.1103/PhysRevB.91.205201}
  {\bibfield  {journal} {\bibinfo  {journal} {Phys. Rev. B}\ }\textbf {\bibinfo
  {volume} {91}},\ \bibinfo {pages} {205201} (\bibinfo {year}
  {2015})}\BibitemShut {NoStop}%
\bibitem [{\citenamefont {Zhao}\ \emph {et~al.}(2016)\citenamefont {Zhao},
  \citenamefont {Chang}, \citenamefont {Tan},\ and\ \citenamefont
  {Kanatzidis}}]{zhao2016snse}%
  \BibitemOpen
  \bibfield  {author} {\bibinfo {author} {\bibfnamefont {L.-D.}\ \bibnamefont
  {Zhao}}, \bibinfo {author} {\bibfnamefont {C.}~\bibnamefont {Chang}},
  \bibinfo {author} {\bibfnamefont {G.}~\bibnamefont {Tan}},\ and\ \bibinfo
  {author} {\bibfnamefont {M.~G.}\ \bibnamefont {Kanatzidis}},\ }\bibfield
  {title} {\bibinfo {title} {{SnSe}: a remarkable new thermoelectric
  material},\ }\href {https://doi.org/10.1039/C6EE01755J} {\bibfield  {journal}
  {\bibinfo  {journal} {Energy \& Environmental Science}\ }\textbf {\bibinfo
  {volume} {9}},\ \bibinfo {pages} {3044} (\bibinfo {year} {2016})}\BibitemShut
  {NoStop}%
\bibitem [{\citenamefont {Hsieh}\ \emph {et~al.}(2012)\citenamefont {Hsieh},
  \citenamefont {Lin}, \citenamefont {Liu}, \citenamefont {Duan}, \citenamefont
  {Bansil},\ and\ \citenamefont {Fu}}]{hsieh2012topological}%
  \BibitemOpen
  \bibfield  {author} {\bibinfo {author} {\bibfnamefont {T.~H.}\ \bibnamefont
  {Hsieh}}, \bibinfo {author} {\bibfnamefont {H.}~\bibnamefont {Lin}}, \bibinfo
  {author} {\bibfnamefont {J.}~\bibnamefont {Liu}}, \bibinfo {author}
  {\bibfnamefont {W.}~\bibnamefont {Duan}}, \bibinfo {author} {\bibfnamefont
  {A.}~\bibnamefont {Bansil}},\ and\ \bibinfo {author} {\bibfnamefont
  {L.}~\bibnamefont {Fu}},\ }\bibfield  {title} {\bibinfo {title} {{Topological
  crystalline insulators in the SnTe material class}},\ }\href
  {https://doi.org/10.1038/ncomms1969} {\bibfield  {journal} {\bibinfo
  {journal} {Nature communications}\ }\textbf {\bibinfo {volume} {3}},\
  \bibinfo {pages} {982} (\bibinfo {year} {2012})}\BibitemShut {NoStop}%
\bibitem [{\citenamefont {Tanaka}\ \emph {et~al.}(2012)\citenamefont {Tanaka},
  \citenamefont {Ren}, \citenamefont {Sato}, \citenamefont {Nakayama},
  \citenamefont {Souma}, \citenamefont {Takahashi}, \citenamefont {Segawa},\
  and\ \citenamefont {Ando}}]{tanaka2012experimental}%
  \BibitemOpen
  \bibfield  {author} {\bibinfo {author} {\bibfnamefont {Y.}~\bibnamefont
  {Tanaka}}, \bibinfo {author} {\bibfnamefont {Z.}~\bibnamefont {Ren}},
  \bibinfo {author} {\bibfnamefont {T.}~\bibnamefont {Sato}}, \bibinfo {author}
  {\bibfnamefont {K.}~\bibnamefont {Nakayama}}, \bibinfo {author}
  {\bibfnamefont {S.}~\bibnamefont {Souma}}, \bibinfo {author} {\bibfnamefont
  {T.}~\bibnamefont {Takahashi}}, \bibinfo {author} {\bibfnamefont
  {K.}~\bibnamefont {Segawa}},\ and\ \bibinfo {author} {\bibfnamefont
  {Y.}~\bibnamefont {Ando}},\ }\bibfield  {title} {\bibinfo {title}
  {{Experimental realization of a topological crystalline insulator in SnTe}},\
  }\href {https://doi.org/10.1038/nphys2442} {\bibfield  {journal} {\bibinfo
  {journal} {Nature Physics}\ }\textbf {\bibinfo {volume} {8}},\ \bibinfo
  {pages} {800} (\bibinfo {year} {2012})}\BibitemShut {NoStop}%
\bibitem [{\citenamefont {Xu}\ \emph {et~al.}(2012)\citenamefont {Xu},
  \citenamefont {Liu}, \citenamefont {Alidoust}, \citenamefont {Neupane},
  \citenamefont {Qian}, \citenamefont {Belopolski}, \citenamefont {Denlinger},
  \citenamefont {Wang}, \citenamefont {Lin}, \citenamefont {Wray} \emph
  {et~al.}}]{xu2012observation}%
  \BibitemOpen
  \bibfield  {author} {\bibinfo {author} {\bibfnamefont {S.-Y.}\ \bibnamefont
  {Xu}}, \bibinfo {author} {\bibfnamefont {C.}~\bibnamefont {Liu}}, \bibinfo
  {author} {\bibfnamefont {N.}~\bibnamefont {Alidoust}}, \bibinfo {author}
  {\bibfnamefont {M.}~\bibnamefont {Neupane}}, \bibinfo {author} {\bibfnamefont
  {D.}~\bibnamefont {Qian}}, \bibinfo {author} {\bibfnamefont {I.}~\bibnamefont
  {Belopolski}}, \bibinfo {author} {\bibfnamefont {J.}~\bibnamefont
  {Denlinger}}, \bibinfo {author} {\bibfnamefont {Y.}~\bibnamefont {Wang}},
  \bibinfo {author} {\bibfnamefont {H.}~\bibnamefont {Lin}}, \bibinfo {author}
  {\bibfnamefont {L.~a.}\ \bibnamefont {Wray}}, \emph {et~al.},\ }\bibfield
  {title} {\bibinfo {title} {Observation of a topological crystalline insulator
  phase and topological phase transition in {Pb$_{1-x}$Sn$_x$ Se$/$Te}},\
  }\href {https://doi.org/10.1038/ncomms2191} {\bibfield  {journal} {\bibinfo
  {journal} {Nature communications}\ }\textbf {\bibinfo {volume} {3}},\
  \bibinfo {pages} {1192} (\bibinfo {year} {2012})}\BibitemShut {NoStop}%
\bibitem [{\citenamefont {Fu}(2011)}]{fu2011topological}%
  \BibitemOpen
  \bibfield  {author} {\bibinfo {author} {\bibfnamefont {L.}~\bibnamefont
  {Fu}},\ }\bibfield  {title} {\bibinfo {title} {Topological crystalline
  insulators},\ }\href {https://doi.org/10.1103/PhysRevLett.106.106802}
  {\bibfield  {journal} {\bibinfo  {journal} {Phys. Rev. Lett.}\ }\textbf
  {\bibinfo {volume} {106}},\ \bibinfo {pages} {106802} (\bibinfo {year}
  {2011})}\BibitemShut {NoStop}%
\bibitem [{\citenamefont {Troullier}\ and\ \citenamefont
  {Martins}(1991)}]{troullier1991efficient}%
  \BibitemOpen
  \bibfield  {author} {\bibinfo {author} {\bibfnamefont {N.}~\bibnamefont
  {Troullier}}\ and\ \bibinfo {author} {\bibfnamefont {J.~L.}\ \bibnamefont
  {Martins}},\ }\bibfield  {title} {\bibinfo {title} {Efficient
  pseudopotentials for plane-wave calculations},\ }\href
  {https://doi.org/10.1103/PhysRevB.43.1993} {\bibfield  {journal} {\bibinfo
  {journal} {Phys. Rev. B}\ }\textbf {\bibinfo {volume} {43}},\ \bibinfo
  {pages} {1993} (\bibinfo {year} {1991})}\BibitemShut {NoStop}%
\bibitem [{\citenamefont {Giannozzi}\ \emph {et~al.}(2009)\citenamefont
  {Giannozzi}, \citenamefont {Baroni}, \citenamefont {Bonini}, \citenamefont
  {Calandra}, \citenamefont {Car}, \citenamefont {Cavazzoni}, \citenamefont
  {Ceresoli}, \citenamefont {Chiarotti}, \citenamefont {Cococcioni},
  \citenamefont {Dabo} \emph {et~al.}}]{giannozzi2009quantum}%
  \BibitemOpen
  \bibfield  {author} {\bibinfo {author} {\bibfnamefont {P.}~\bibnamefont
  {Giannozzi}}, \bibinfo {author} {\bibfnamefont {S.}~\bibnamefont {Baroni}},
  \bibinfo {author} {\bibfnamefont {N.}~\bibnamefont {Bonini}}, \bibinfo
  {author} {\bibfnamefont {M.}~\bibnamefont {Calandra}}, \bibinfo {author}
  {\bibfnamefont {R.}~\bibnamefont {Car}}, \bibinfo {author} {\bibfnamefont
  {C.}~\bibnamefont {Cavazzoni}}, \bibinfo {author} {\bibfnamefont
  {D.}~\bibnamefont {Ceresoli}}, \bibinfo {author} {\bibfnamefont {G.~L.}\
  \bibnamefont {Chiarotti}}, \bibinfo {author} {\bibfnamefont {M.}~\bibnamefont
  {Cococcioni}}, \bibinfo {author} {\bibfnamefont {I.}~\bibnamefont {Dabo}},
  \emph {et~al.},\ }\bibfield  {title} {\bibinfo {title} {{QUANTUM ESPRESSO: a
  modular and open-source software project for quantum simulations of
  materials}},\ }\href {https://doi.org/10.1088/0953-8984/21/39/395502}
  {\bibfield  {journal} {\bibinfo  {journal} {Journal of physics: Condensed
  matter}\ }\textbf {\bibinfo {volume} {21}},\ \bibinfo {pages} {395502}
  (\bibinfo {year} {2009})}\BibitemShut {NoStop}%
\bibitem [{\citenamefont {Giannozzi}\ \emph {et~al.}(2017)\citenamefont
  {Giannozzi}, \citenamefont {Andreussi}, \citenamefont {Brumme}, \citenamefont
  {Bunau}, \citenamefont {Nardelli}, \citenamefont {Calandra}, \citenamefont
  {Car}, \citenamefont {Cavazzoni}, \citenamefont {Ceresoli}, \citenamefont
  {Cococcioni} \emph {et~al.}}]{giannozzi2017advanced}%
  \BibitemOpen
  \bibfield  {author} {\bibinfo {author} {\bibfnamefont {P.}~\bibnamefont
  {Giannozzi}}, \bibinfo {author} {\bibfnamefont {O.}~\bibnamefont
  {Andreussi}}, \bibinfo {author} {\bibfnamefont {T.}~\bibnamefont {Brumme}},
  \bibinfo {author} {\bibfnamefont {O.}~\bibnamefont {Bunau}}, \bibinfo
  {author} {\bibfnamefont {M.~B.}\ \bibnamefont {Nardelli}}, \bibinfo {author}
  {\bibfnamefont {M.}~\bibnamefont {Calandra}}, \bibinfo {author}
  {\bibfnamefont {R.}~\bibnamefont {Car}}, \bibinfo {author} {\bibfnamefont
  {C.}~\bibnamefont {Cavazzoni}}, \bibinfo {author} {\bibfnamefont
  {D.}~\bibnamefont {Ceresoli}}, \bibinfo {author} {\bibfnamefont
  {M.}~\bibnamefont {Cococcioni}}, \emph {et~al.},\ }\bibfield  {title}
  {\bibinfo {title} {Advanced capabilities for materials modelling with quantum
  espresso},\ }\href {https://doi.org/10.1088/1361-648X/aa8f79} {\bibfield
  {journal} {\bibinfo  {journal} {Journal of physics: Condensed matter}\
  }\textbf {\bibinfo {volume} {29}},\ \bibinfo {pages} {465901} (\bibinfo
  {year} {2017})}\BibitemShut {NoStop}%
\bibitem [{\citenamefont {Giannozzi}\ \emph {et~al.}(2020)\citenamefont
  {Giannozzi}, \citenamefont {Baseggio}, \citenamefont {Bonf{\`a}},
  \citenamefont {Brunato}, \citenamefont {Car}, \citenamefont {Carnimeo},
  \citenamefont {Cavazzoni}, \citenamefont {De~Gironcoli}, \citenamefont
  {Delugas}, \citenamefont {Ferrari~Ruffino} \emph
  {et~al.}}]{giannozzi2020quantum}%
  \BibitemOpen
  \bibfield  {author} {\bibinfo {author} {\bibfnamefont {P.}~\bibnamefont
  {Giannozzi}}, \bibinfo {author} {\bibfnamefont {O.}~\bibnamefont {Baseggio}},
  \bibinfo {author} {\bibfnamefont {P.}~\bibnamefont {Bonf{\`a}}}, \bibinfo
  {author} {\bibfnamefont {D.}~\bibnamefont {Brunato}}, \bibinfo {author}
  {\bibfnamefont {R.}~\bibnamefont {Car}}, \bibinfo {author} {\bibfnamefont
  {I.}~\bibnamefont {Carnimeo}}, \bibinfo {author} {\bibfnamefont
  {C.}~\bibnamefont {Cavazzoni}}, \bibinfo {author} {\bibfnamefont
  {S.}~\bibnamefont {De~Gironcoli}}, \bibinfo {author} {\bibfnamefont
  {P.}~\bibnamefont {Delugas}}, \bibinfo {author} {\bibfnamefont
  {F.}~\bibnamefont {Ferrari~Ruffino}}, \emph {et~al.},\ }\bibfield  {title}
  {\bibinfo {title} {Quantum espresso toward the exascale},\ }\href
  {https://doi.org/10.1063/5.0005082} {\bibfield  {journal} {\bibinfo
  {journal} {The Journal of chemical physics}\ }\textbf {\bibinfo {volume}
  {152}},\ \bibinfo {pages} {154105} (\bibinfo {year} {2020})}\BibitemShut
  {NoStop}%
\bibitem [{\citenamefont {Marzari}\ and\ \citenamefont
  {Vanderbilt}(1997)}]{marzari1997maximally}%
  \BibitemOpen
  \bibfield  {author} {\bibinfo {author} {\bibfnamefont {N.}~\bibnamefont
  {Marzari}}\ and\ \bibinfo {author} {\bibfnamefont {D.}~\bibnamefont
  {Vanderbilt}},\ }\bibfield  {title} {\bibinfo {title} {Maximally localized
  generalized wannier functions for composite energy bands},\ }\href
  {https://doi.org/10.1103/PhysRevB.56.12847} {\bibfield  {journal} {\bibinfo
  {journal} {Phys. Rev. B}\ }\textbf {\bibinfo {volume} {56}},\ \bibinfo
  {pages} {12847} (\bibinfo {year} {1997})}\BibitemShut {NoStop}%
\bibitem [{\citenamefont {Souza}\ \emph {et~al.}(2001)\citenamefont {Souza},
  \citenamefont {Marzari},\ and\ \citenamefont
  {Vanderbilt}}]{souza2001maximally}%
  \BibitemOpen
  \bibfield  {author} {\bibinfo {author} {\bibfnamefont {I.}~\bibnamefont
  {Souza}}, \bibinfo {author} {\bibfnamefont {N.}~\bibnamefont {Marzari}},\
  and\ \bibinfo {author} {\bibfnamefont {D.}~\bibnamefont {Vanderbilt}},\
  }\bibfield  {title} {\bibinfo {title} {Maximally localized wannier functions
  for entangled energy bands},\ }\href
  {https://doi.org/10.1103/PhysRevB.65.035109} {\bibfield  {journal} {\bibinfo
  {journal} {Phys. Rev. B}\ }\textbf {\bibinfo {volume} {65}},\ \bibinfo
  {pages} {035109} (\bibinfo {year} {2001})}\BibitemShut {NoStop}%
\bibitem [{\citenamefont {Mostofi}\ \emph {et~al.}(2008)\citenamefont
  {Mostofi}, \citenamefont {Yates}, \citenamefont {Lee}, \citenamefont {Souza},
  \citenamefont {Vanderbilt},\ and\ \citenamefont
  {Marzari}}]{mostofi2008wannier90}%
  \BibitemOpen
  \bibfield  {author} {\bibinfo {author} {\bibfnamefont {A.~A.}\ \bibnamefont
  {Mostofi}}, \bibinfo {author} {\bibfnamefont {J.~R.}\ \bibnamefont {Yates}},
  \bibinfo {author} {\bibfnamefont {Y.-S.}\ \bibnamefont {Lee}}, \bibinfo
  {author} {\bibfnamefont {I.}~\bibnamefont {Souza}}, \bibinfo {author}
  {\bibfnamefont {D.}~\bibnamefont {Vanderbilt}},\ and\ \bibinfo {author}
  {\bibfnamefont {N.}~\bibnamefont {Marzari}},\ }\bibfield  {title} {\bibinfo
  {title} {wannier90: A tool for obtaining maximally-localised wannier
  functions},\ }\href {https://doi.org/10.1016/j.cpc.2007.11.016} {\bibfield
  {journal} {\bibinfo  {journal} {Computer physics communications}\ }\textbf
  {\bibinfo {volume} {178}},\ \bibinfo {pages} {685} (\bibinfo {year}
  {2008})}\BibitemShut {NoStop}%
\bibitem [{\citenamefont {Giustino}\ \emph {et~al.}(2007)\citenamefont
  {Giustino}, \citenamefont {Cohen},\ and\ \citenamefont
  {Louie}}]{giustino2007electron}%
  \BibitemOpen
  \bibfield  {author} {\bibinfo {author} {\bibfnamefont {F.}~\bibnamefont
  {Giustino}}, \bibinfo {author} {\bibfnamefont {M.~L.}\ \bibnamefont
  {Cohen}},\ and\ \bibinfo {author} {\bibfnamefont {S.~G.}\ \bibnamefont
  {Louie}},\ }\bibfield  {title} {\bibinfo {title} {{Electron-phonon
  interaction using Wannier functions}},\ }\href
  {https://doi.org/10.1103/PhysRevB.76.165108} {\bibfield  {journal} {\bibinfo
  {journal} {Phys. Rev. B}\ }\textbf {\bibinfo {volume} {76}},\ \bibinfo
  {pages} {165108} (\bibinfo {year} {2007})}\BibitemShut {NoStop}%
\bibitem [{\citenamefont {Ponc{\'e}}\ \emph {et~al.}(2016)\citenamefont
  {Ponc{\'e}}, \citenamefont {Margine}, \citenamefont {Verdi},\ and\
  \citenamefont {Giustino}}]{ponce2016epw}%
  \BibitemOpen
  \bibfield  {author} {\bibinfo {author} {\bibfnamefont {S.}~\bibnamefont
  {Ponc{\'e}}}, \bibinfo {author} {\bibfnamefont {E.~R.}\ \bibnamefont
  {Margine}}, \bibinfo {author} {\bibfnamefont {C.}~\bibnamefont {Verdi}},\
  and\ \bibinfo {author} {\bibfnamefont {F.}~\bibnamefont {Giustino}},\
  }\bibfield  {title} {\bibinfo {title} {{EPW: Electron--phonon coupling,
  transport and superconducting properties using maximally localized Wannier
  functions}},\ }\href {https://doi.org/10.1016/j.cpc.2016.07.028} {\bibfield
  {journal} {\bibinfo  {journal} {Computer Physics Communications}\ }\textbf
  {\bibinfo {volume} {209}},\ \bibinfo {pages} {116} (\bibinfo {year}
  {2016})}\BibitemShut {NoStop}%
\bibitem [{\citenamefont {Margine}\ and\ \citenamefont
  {Giustino}(2013)}]{margine2013anisotropic}%
  \BibitemOpen
  \bibfield  {author} {\bibinfo {author} {\bibfnamefont {E.~R.}\ \bibnamefont
  {Margine}}\ and\ \bibinfo {author} {\bibfnamefont {F.}~\bibnamefont
  {Giustino}},\ }\bibfield  {title} {\bibinfo {title} {{Anisotropic
  Migdal-Eliashberg theory using Wannier functions}},\ }\href
  {https://doi.org/10.1103/PhysRevB.87.024505} {\bibfield  {journal} {\bibinfo
  {journal} {Phys. Rev. B}\ }\textbf {\bibinfo {volume} {87}},\ \bibinfo
  {pages} {024505} (\bibinfo {year} {2013})}\BibitemShut {NoStop}%
\bibitem [{\citenamefont {Allen}\ and\ \citenamefont
  {Mitrovi{\'c}}(1983)}]{allen1983theory}%
  \BibitemOpen
  \bibfield  {author} {\bibinfo {author} {\bibfnamefont {P.~B.}\ \bibnamefont
  {Allen}}\ and\ \bibinfo {author} {\bibfnamefont {B.}~\bibnamefont
  {Mitrovi{\'c}}},\ }\bibfield  {title} {\bibinfo {title} {Theory of
  superconducting tc},\ }\href {https://doi.org/10.1016/S0081-1947(08)60665-7}
  {\bibfield  {journal} {\bibinfo  {journal} {Solid state physics}\ }\textbf
  {\bibinfo {volume} {37}},\ \bibinfo {pages} {1} (\bibinfo {year}
  {1983})}\BibitemShut {NoStop}%
\bibitem [{\citenamefont {Allen}\ and\ \citenamefont
  {Dynes}(1975)}]{allen1975transition}%
  \BibitemOpen
  \bibfield  {author} {\bibinfo {author} {\bibfnamefont {P.~B.}\ \bibnamefont
  {Allen}}\ and\ \bibinfo {author} {\bibfnamefont {R.~C.}\ \bibnamefont
  {Dynes}},\ }\bibfield  {title} {\bibinfo {title} {Transition temperature of
  strong-coupled superconductors reanalyzed},\ }\href
  {https://doi.org/10.1103/PhysRevB.12.905} {\bibfield  {journal} {\bibinfo
  {journal} {Phys. Rev. B}\ }\textbf {\bibinfo {volume} {12}},\ \bibinfo
  {pages} {905} (\bibinfo {year} {1975})}\BibitemShut {NoStop}%
\bibitem [{\citenamefont {McMillan}(1968)}]{mcmillan1968transition}%
  \BibitemOpen
  \bibfield  {author} {\bibinfo {author} {\bibfnamefont {W.~L.}\ \bibnamefont
  {McMillan}},\ }\bibfield  {title} {\bibinfo {title} {{Transition Temperature
  of Strong-Coupled Superconductors}},\ }\href
  {https://doi.org/10.1103/PhysRev.167.331} {\bibfield  {journal} {\bibinfo
  {journal} {Phys. Rev.}\ }\textbf {\bibinfo {volume} {167}},\ \bibinfo {pages}
  {331} (\bibinfo {year} {1968})}\BibitemShut {NoStop}%
\bibitem [{\citenamefont {Henkelman}\ \emph {et~al.}(2006)\citenamefont
  {Henkelman}, \citenamefont {Arnaldsson},\ and\ \citenamefont
  {J{\'o}nsson}}]{henkelman2006fast}%
  \BibitemOpen
  \bibfield  {author} {\bibinfo {author} {\bibfnamefont {G.}~\bibnamefont
  {Henkelman}}, \bibinfo {author} {\bibfnamefont {A.}~\bibnamefont
  {Arnaldsson}},\ and\ \bibinfo {author} {\bibfnamefont {H.}~\bibnamefont
  {J{\'o}nsson}},\ }\bibfield  {title} {\bibinfo {title} {A fast and robust
  algorithm for bader decomposition of charge density},\ }\href
  {https://doi.org/10.1016/j.commatsci.2005.04.010} {\bibfield  {journal}
  {\bibinfo  {journal} {Computational Materials Science}\ }\textbf {\bibinfo
  {volume} {36}},\ \bibinfo {pages} {354} (\bibinfo {year} {2006})}\BibitemShut
  {NoStop}%
\bibitem [{\citenamefont {Chong}\ \emph {et~al.}(2008)\citenamefont {Chong},
  \citenamefont {Wen},\ and\ \citenamefont
  {Solja{\v{c}}i{\'c}}}]{chong2008effective}%
  \BibitemOpen
  \bibfield  {author} {\bibinfo {author} {\bibfnamefont {Y.~D.}\ \bibnamefont
  {Chong}}, \bibinfo {author} {\bibfnamefont {X.-G.}\ \bibnamefont {Wen}},\
  and\ \bibinfo {author} {\bibfnamefont {M.}~\bibnamefont
  {Solja{\v{c}}i{\'c}}},\ }\bibfield  {title} {\bibinfo {title} {Effective
  theory of quadratic degeneracies},\ }\href
  {https://doi.org/10.1103/PhysRevB.77.235125} {\bibfield  {journal} {\bibinfo
  {journal} {Phys. Rev. B}\ }\textbf {\bibinfo {volume} {77}},\ \bibinfo
  {pages} {235125} (\bibinfo {year} {2008})}\BibitemShut {NoStop}%
\bibitem [{\citenamefont {Sun}\ \emph {et~al.}(2009)\citenamefont {Sun},
  \citenamefont {Yao}, \citenamefont {Fradkin},\ and\ \citenamefont
  {Kivelson}}]{sun2009topological}%
  \BibitemOpen
  \bibfield  {author} {\bibinfo {author} {\bibfnamefont {K.}~\bibnamefont
  {Sun}}, \bibinfo {author} {\bibfnamefont {H.}~\bibnamefont {Yao}}, \bibinfo
  {author} {\bibfnamefont {E.}~\bibnamefont {Fradkin}},\ and\ \bibinfo {author}
  {\bibfnamefont {S.~A.}\ \bibnamefont {Kivelson}},\ }\bibfield  {title}
  {\bibinfo {title} {{Topological Insulators and Nematic Phases from
  Spontaneous Symmetry Breaking in 2D Fermi Systems with a Quadratic Band
  Crossing}},\ }\href {https://doi.org/10.1103/PhysRevLett.103.046811}
  {\bibfield  {journal} {\bibinfo  {journal} {Phys. Rev. Lett.}\ }\textbf
  {\bibinfo {volume} {103}},\ \bibinfo {pages} {046811} (\bibinfo {year}
  {2009})}\BibitemShut {NoStop}%
\bibitem [{\citenamefont {Barone}\ \emph {et~al.}(2013)\citenamefont {Barone},
  \citenamefont {Rauch}, \citenamefont {Di~Sante}, \citenamefont {Henk},
  \citenamefont {Mertig},\ and\ \citenamefont {Picozzi}}]{barone2013pressure}%
  \BibitemOpen
  \bibfield  {author} {\bibinfo {author} {\bibfnamefont {P.}~\bibnamefont
  {Barone}}, \bibinfo {author} {\bibfnamefont {T.}~\bibnamefont {Rauch}},
  \bibinfo {author} {\bibfnamefont {D.}~\bibnamefont {Di~Sante}}, \bibinfo
  {author} {\bibfnamefont {J.}~\bibnamefont {Henk}}, \bibinfo {author}
  {\bibfnamefont {I.}~\bibnamefont {Mertig}},\ and\ \bibinfo {author}
  {\bibfnamefont {S.}~\bibnamefont {Picozzi}},\ }\bibfield  {title} {\bibinfo
  {title} {Pressure-induced topological phase transitions in rocksalt
  chalcogenides},\ }\href {https://doi.org/10.1103/PhysRevB.88.045207}
  {\bibfield  {journal} {\bibinfo  {journal} {Phys. Rev. B}\ }\textbf {\bibinfo
  {volume} {88}},\ \bibinfo {pages} {045207} (\bibinfo {year}
  {2013})}\BibitemShut {NoStop}%
\bibitem [{\citenamefont {Chen}\ \emph {et~al.}(2015)\citenamefont {Chen},
  \citenamefont {Lu},\ and\ \citenamefont {Kee}}]{chen2015topological}%
  \BibitemOpen
  \bibfield  {author} {\bibinfo {author} {\bibfnamefont {Y.}~\bibnamefont
  {Chen}}, \bibinfo {author} {\bibfnamefont {Y.-M.}\ \bibnamefont {Lu}},\ and\
  \bibinfo {author} {\bibfnamefont {H.-Y.}\ \bibnamefont {Kee}},\ }\bibfield
  {title} {\bibinfo {title} {Topological crystalline metal in orthorhombic
  perovskite iridates},\ }\href {https://doi.org/10.1038/ncomms7593} {\bibfield
   {journal} {\bibinfo  {journal} {Nature communications}\ }\textbf {\bibinfo
  {volume} {6}},\ \bibinfo {pages} {6593} (\bibinfo {year} {2015})}\BibitemShut
  {NoStop}%
\bibitem [{\citenamefont {Kim}\ \emph {et~al.}(2015)\citenamefont {Kim},
  \citenamefont {Chen},\ and\ \citenamefont {Kee}}]{kim2015surface}%
  \BibitemOpen
  \bibfield  {author} {\bibinfo {author} {\bibfnamefont {H.-S.}\ \bibnamefont
  {Kim}}, \bibinfo {author} {\bibfnamefont {Y.}~\bibnamefont {Chen}},\ and\
  \bibinfo {author} {\bibfnamefont {H.-Y.}\ \bibnamefont {Kee}},\ }\bibfield
  {title} {\bibinfo {title} {Surface states of perovskite iridates
  ${\mathrm{airo}}_{3}$: Signatures of a topological crystalline metal with
  nontrivial ${\mathbb{z}}_{2}$ index},\ }\href
  {https://doi.org/10.1103/PhysRevB.91.235103} {\bibfield  {journal} {\bibinfo
  {journal} {Phys. Rev. B}\ }\textbf {\bibinfo {volume} {91}},\ \bibinfo
  {pages} {235103} (\bibinfo {year} {2015})}\BibitemShut {NoStop}%
\bibitem [{\citenamefont {Berry}(1984)}]{berry1984quantal}%
  \BibitemOpen
  \bibfield  {author} {\bibinfo {author} {\bibfnamefont {M.~V.}\ \bibnamefont
  {Berry}},\ }\bibfield  {title} {\bibinfo {title} {Quantal phase factors
  accompanying adiabatic changes},\ }\href
  {https://doi.org/10.1098/rspa.1984.0023} {\bibfield  {journal} {\bibinfo
  {journal} {Proceedings of the Royal Society of London. A. Mathematical and
  Physical Sciences}\ }\textbf {\bibinfo {volume} {392}},\ \bibinfo {pages}
  {45} (\bibinfo {year} {1984})}\BibitemShut {NoStop}%
\bibitem [{\citenamefont {Xiao}\ \emph {et~al.}(2010)\citenamefont {Xiao},
  \citenamefont {Chang},\ and\ \citenamefont {Niu}}]{xiao2010berry}%
  \BibitemOpen
  \bibfield  {author} {\bibinfo {author} {\bibfnamefont {D.}~\bibnamefont
  {Xiao}}, \bibinfo {author} {\bibfnamefont {M.-C.}\ \bibnamefont {Chang}},\
  and\ \bibinfo {author} {\bibfnamefont {Q.}~\bibnamefont {Niu}},\ }\bibfield
  {title} {\bibinfo {title} {Berry phase effects on electronic properties},\
  }\href {https://doi.org/10.1103/RevModPhys.82.1959} {\bibfield  {journal}
  {\bibinfo  {journal} {Rev. Mod. Phys.}\ }\textbf {\bibinfo {volume} {82}},\
  \bibinfo {pages} {1959} (\bibinfo {year} {2010})}\BibitemShut {NoStop}%
\bibitem [{\citenamefont {Haldane}(1988)}]{haldane1988model}%
  \BibitemOpen
  \bibfield  {author} {\bibinfo {author} {\bibfnamefont {F.~D.~M.}\
  \bibnamefont {Haldane}},\ }\bibfield  {title} {\bibinfo {title} {Model for a
  quantum hall effect without landau levels: Condensed-matter realization of
  the "parity anomaly"},\ }\href {https://doi.org/10.1103/PhysRevLett.61.2015}
  {\bibfield  {journal} {\bibinfo  {journal} {Phys. Rev. Lett.}\ }\textbf
  {\bibinfo {volume} {61}},\ \bibinfo {pages} {2015} (\bibinfo {year}
  {1988})}\BibitemShut {NoStop}%
\bibitem [{\citenamefont {Teo}\ \emph {et~al.}(2008)\citenamefont {Teo},
  \citenamefont {Fu},\ and\ \citenamefont {Kane}}]{teo2008surface}%
  \BibitemOpen
  \bibfield  {author} {\bibinfo {author} {\bibfnamefont {J.~C.~Y.}\
  \bibnamefont {Teo}}, \bibinfo {author} {\bibfnamefont {L.}~\bibnamefont
  {Fu}},\ and\ \bibinfo {author} {\bibfnamefont {C.~L.}\ \bibnamefont {Kane}},\
  }\bibfield  {title} {\bibinfo {title} {Surface states and topological
  invariants in three-dimensional topological insulators: Application to
  ${\text{bi}}_{1\ensuremath{-}x}{\text{sb}}_{x}$},\ }\href
  {https://doi.org/10.1103/PhysRevB.78.045426} {\bibfield  {journal} {\bibinfo
  {journal} {Phys. Rev. B}\ }\textbf {\bibinfo {volume} {78}},\ \bibinfo
  {pages} {045426} (\bibinfo {year} {2008})}\BibitemShut {NoStop}%
\bibitem [{\citenamefont {Liu}\ \emph {et~al.}(2013)\citenamefont {Liu},
  \citenamefont {Duan},\ and\ \citenamefont {Fu}}]{junwei2013two}%
  \BibitemOpen
  \bibfield  {author} {\bibinfo {author} {\bibfnamefont {J.}~\bibnamefont
  {Liu}}, \bibinfo {author} {\bibfnamefont {W.}~\bibnamefont {Duan}},\ and\
  \bibinfo {author} {\bibfnamefont {L.}~\bibnamefont {Fu}},\ }\bibfield
  {title} {\bibinfo {title} {Two types of surface states in topological
  crystalline insulators},\ }\href {https://doi.org/10.1103/PhysRevB.88.241303}
  {\bibfield  {journal} {\bibinfo  {journal} {Phys. Rev. B}\ }\textbf {\bibinfo
  {volume} {88}},\ \bibinfo {pages} {241303} (\bibinfo {year}
  {2013})}\BibitemShut {NoStop}%
\bibitem [{\citenamefont {Alexandradinata}\ and\ \citenamefont
  {Bernevig}(2016)}]{alexandradinata2016berry}%
  \BibitemOpen
  \bibfield  {author} {\bibinfo {author} {\bibfnamefont {A.}~\bibnamefont
  {Alexandradinata}}\ and\ \bibinfo {author} {\bibfnamefont {B.~A.}\
  \bibnamefont {Bernevig}},\ }\bibfield  {title} {\bibinfo {title} {Berry-phase
  description of topological crystalline insulators},\ }\href
  {https://doi.org/10.1103/PhysRevB.93.205104} {\bibfield  {journal} {\bibinfo
  {journal} {Phys. Rev. B}\ }\textbf {\bibinfo {volume} {93}},\ \bibinfo
  {pages} {205104} (\bibinfo {year} {2016})}\BibitemShut {NoStop}%
\bibitem [{\citenamefont {Wu}\ \emph {et~al.}(2018)\citenamefont {Wu},
  \citenamefont {Zhang}, \citenamefont {Song}, \citenamefont {Troyer},\ and\
  \citenamefont {Soluyanov}}]{wu2017wanniertools}%
  \BibitemOpen
  \bibfield  {author} {\bibinfo {author} {\bibfnamefont {Q.}~\bibnamefont
  {Wu}}, \bibinfo {author} {\bibfnamefont {S.}~\bibnamefont {Zhang}}, \bibinfo
  {author} {\bibfnamefont {H.-F.}\ \bibnamefont {Song}}, \bibinfo {author}
  {\bibfnamefont {M.}~\bibnamefont {Troyer}},\ and\ \bibinfo {author}
  {\bibfnamefont {A.~A.}\ \bibnamefont {Soluyanov}},\ }\bibfield  {title}
  {\bibinfo {title} {Wanniertools : An open-source software package for novel
  topological materials},\ }\href
  {https://doi.org/https://doi.org/10.1016/j.cpc.2017.09.033} {\bibfield
  {journal} {\bibinfo  {journal} {Computer Physics Communications}\ }\textbf
  {\bibinfo {volume} {224}},\ \bibinfo {pages} {405 } (\bibinfo {year}
  {2018})}\BibitemShut {NoStop}%
\bibitem [{\citenamefont {Schindler}\ \emph {et~al.}(2018)\citenamefont
  {Schindler}, \citenamefont {Cook}, \citenamefont {Vergniory}, \citenamefont
  {Wang}, \citenamefont {Parkin}, \citenamefont {Bernevig},\ and\ \citenamefont
  {Neupert}}]{schindler2018higher}%
  \BibitemOpen
  \bibfield  {author} {\bibinfo {author} {\bibfnamefont {F.}~\bibnamefont
  {Schindler}}, \bibinfo {author} {\bibfnamefont {A.~M.}\ \bibnamefont {Cook}},
  \bibinfo {author} {\bibfnamefont {M.~G.}\ \bibnamefont {Vergniory}}, \bibinfo
  {author} {\bibfnamefont {Z.}~\bibnamefont {Wang}}, \bibinfo {author}
  {\bibfnamefont {S.~S.}\ \bibnamefont {Parkin}}, \bibinfo {author}
  {\bibfnamefont {B.~A.}\ \bibnamefont {Bernevig}},\ and\ \bibinfo {author}
  {\bibfnamefont {T.}~\bibnamefont {Neupert}},\ }\bibfield  {title} {\bibinfo
  {title} {Higher-order topological insulators},\ }\href
  {https://doi.org/10.1126/sciadv.aat0346} {\bibfield  {journal} {\bibinfo
  {journal} {Science Advances}\ }\textbf {\bibinfo {volume} {4}},\ \bibinfo
  {pages} {eaat0346} (\bibinfo {year} {2018})}\BibitemShut {NoStop}%
\bibitem [{\citenamefont {Wang}\ \emph {et~al.}(2013)\citenamefont {Wang},
  \citenamefont {Tsai}, \citenamefont {Lin}, \citenamefont {Xu}, \citenamefont
  {Neupane}, \citenamefont {Hasan},\ and\ \citenamefont
  {Bansil}}]{wang2013nontrivial}%
  \BibitemOpen
  \bibfield  {author} {\bibinfo {author} {\bibfnamefont {Y.~J.}\ \bibnamefont
  {Wang}}, \bibinfo {author} {\bibfnamefont {W.-F.}\ \bibnamefont {Tsai}},
  \bibinfo {author} {\bibfnamefont {H.}~\bibnamefont {Lin}}, \bibinfo {author}
  {\bibfnamefont {S.-Y.}\ \bibnamefont {Xu}}, \bibinfo {author} {\bibfnamefont
  {M.}~\bibnamefont {Neupane}}, \bibinfo {author} {\bibfnamefont {M.~Z.}\
  \bibnamefont {Hasan}},\ and\ \bibinfo {author} {\bibfnamefont
  {A.}~\bibnamefont {Bansil}},\ }\bibfield  {title} {\bibinfo {title}
  {Nontrivial spin texture of the coaxial dirac cones on the surface of
  topological crystalline insulator snte},\ }\href
  {https://doi.org/10.1103/PhysRevB.87.235317} {\bibfield  {journal} {\bibinfo
  {journal} {Phys. Rev. B}\ }\textbf {\bibinfo {volume} {87}},\ \bibinfo
  {pages} {235317} (\bibinfo {year} {2013})}\BibitemShut {NoStop}%
\bibitem [{\citenamefont {Liu}\ and\ \citenamefont
  {Fu}(2015)}]{liu2015electrically}%
  \BibitemOpen
  \bibfield  {author} {\bibinfo {author} {\bibfnamefont {J.}~\bibnamefont
  {Liu}}\ and\ \bibinfo {author} {\bibfnamefont {L.}~\bibnamefont {Fu}},\
  }\bibfield  {title} {\bibinfo {title} {Electrically tunable quantum spin hall
  state in topological crystalline insulator thin films},\ }\href
  {https://doi.org/10.1103/PhysRevB.91.081407} {\bibfield  {journal} {\bibinfo
  {journal} {Phys. Rev. B}\ }\textbf {\bibinfo {volume} {91}},\ \bibinfo
  {pages} {081407} (\bibinfo {year} {2015})}\BibitemShut {NoStop}%
\bibitem [{\citenamefont {Safaei}\ \emph {et~al.}(2015)\citenamefont {Safaei},
  \citenamefont {Galicka}, \citenamefont {Kacman},\ and\ \citenamefont
  {Buczko}}]{safaei2015quantum}%
  \BibitemOpen
  \bibfield  {author} {\bibinfo {author} {\bibfnamefont {S.}~\bibnamefont
  {Safaei}}, \bibinfo {author} {\bibfnamefont {M.}~\bibnamefont {Galicka}},
  \bibinfo {author} {\bibfnamefont {P.}~\bibnamefont {Kacman}},\ and\ \bibinfo
  {author} {\bibfnamefont {R.}~\bibnamefont {Buczko}},\ }\bibfield  {title}
  {\bibinfo {title} {Quantum spin hall effect in iv-vi topological crystalline
  insulators},\ }\href {https://doi.org/10.1088/1367-2630/17/6/063041}
  {\bibfield  {journal} {\bibinfo  {journal} {New Journal of Physics}\ }\textbf
  {\bibinfo {volume} {17}},\ \bibinfo {pages} {063041} (\bibinfo {year}
  {2015})}\BibitemShut {NoStop}%
\bibitem [{\citenamefont {Matsuura}\ \emph {et~al.}(2022)\citenamefont
  {Matsuura}, \citenamefont {Mukuda},\ and\ \citenamefont
  {Miyake}}]{matsuura2022valence}%
  \BibitemOpen
  \bibfield  {author} {\bibinfo {author} {\bibfnamefont {H.}~\bibnamefont
  {Matsuura}}, \bibinfo {author} {\bibfnamefont {H.}~\bibnamefont {Mukuda}},\
  and\ \bibinfo {author} {\bibfnamefont {K.}~\bibnamefont {Miyake}},\
  }\bibfield  {title} {\bibinfo {title} {Valence skipping phenomena, charge
  kondo effect, and superconductivity},\ }\href
  {https://doi.org/10.1007/s43673-022-00056-1} {\bibfield  {journal} {\bibinfo
  {journal} {AAPPS Bulletin}\ }\textbf {\bibinfo {volume} {32}},\ \bibinfo
  {pages} {30} (\bibinfo {year} {2022})}\BibitemShut {NoStop}%
\bibitem [{\citenamefont {Patel}\ \emph {et~al.}(2024)\citenamefont {Patel},
  \citenamefont {Jena},\ and\ \citenamefont {Taraphder}}]{patel2024electron}%
  \BibitemOpen
  \bibfield  {author} {\bibinfo {author} {\bibfnamefont {S.}~\bibnamefont
  {Patel}}, \bibinfo {author} {\bibfnamefont {S.}~\bibnamefont {Jena}},\ and\
  \bibinfo {author} {\bibfnamefont {A.}~\bibnamefont {Taraphder}},\ }\bibfield
  {title} {\bibinfo {title} {Electron-phonon coupling, critical temperatures,
  and gaps in ${\mathrm{nbse}}_{2}/{\mathrm{mos}}_{2}$ ising superconductors},\
  }\href {https://doi.org/10.1103/PhysRevB.110.014507} {\bibfield  {journal}
  {\bibinfo  {journal} {Phys. Rev. B}\ }\textbf {\bibinfo {volume} {110}},\
  \bibinfo {pages} {014507} (\bibinfo {year} {2024})}\BibitemShut {NoStop}%
\bibitem [{\citenamefont {Dzero}\ and\ \citenamefont
  {Schmalian}(2005)}]{dzero2005superconductivity}%
  \BibitemOpen
  \bibfield  {author} {\bibinfo {author} {\bibfnamefont {M.}~\bibnamefont
  {Dzero}}\ and\ \bibinfo {author} {\bibfnamefont {J.}~\bibnamefont
  {Schmalian}},\ }\bibfield  {title} {\bibinfo {title} {Superconductivity in
  charge kondo systems},\ }\href
  {https://doi.org/10.1103/PhysRevLett.94.157003} {\bibfield  {journal}
  {\bibinfo  {journal} {Phys. Rev. Lett.}\ }\textbf {\bibinfo {volume} {94}},\
  \bibinfo {pages} {157003} (\bibinfo {year} {2005})}\BibitemShut {NoStop}%
\end{thebibliography}%
	
	%\begin{thebibliography}{99}

	%%%%%%%%%%%%%%%%%%%%%%%%%%%%%%%%%%%%%%%%%%%%%%%%%%%%%%%%%%%%%%%%%%%%%%%%%%%%%%%%%%%%%%%%%%%%%%%%%%%%%%%%%%%%%%%%%%%%
	%\end{thebibliography}
	%-------------------------------------------------------------------------------------------------------------------------
	
\end{document}